# Formation of Thiocarbonic Acid ($H_2CS_3$) – the Sulfur Counterpart of Carbonic Acid ($H_2CO_3$) – in Interstellar Analog Ices

Lina Coulaud,[1,2†] Jia Wang,[1,2] Ashanie Herath,[1,2] Andrew M. Turner,[1,2] Mason Mcanally,[1,2] Ryan C. Fortenberry,[3*] Ralf I. Kaiser[1,2*]

[1] *W. M. Keck Research Laboratory in Astrochemistry, University of Hawaii at Manoa, Honolulu, HI 96822, USA*

[2] *Department of Chemistry, University of Hawaii at Manoa, Honolulu, HI 96822, USA*

[3] *Department of Chemistry & Biochemistry, University of Mississippi, MS 38677, USA*

[†] *Present address: Département de Chimie, Université Paris-Saclay, Gif-sur-Yvette, 91190, France*

[*]Corresponding Author: r410@olemiss.edu, ralfk@hawaii.edu

## Abstract

The first experimental formation of thiocarbonic acid ($H_2CS_3$) is presented in this work from low-temperature interstellar ice analogs composed of hydrogen sulfide ($H_2S$) and carbon disulfide ($CS_2$), exposed to electron irradiation simulating the impact of galactic cosmic rays (GCRs) on interstellar ices. The recent attention brought to sulfur-bearing molecules as well as the recent detection of carbonic acid ($H_2CO_3$) in the interstellar medium (ISM) invites study of the interstellar detection of the sulfur counterpart, thiocarbonic acid. However, the interstellar formation pathways of thiocarbonic acid have remained elusive. In this work, thiocarbonic acid was identified in the gas phase during the temperature programmed desorption (TPD) using isomer-selective single photoionization reflectron time-of-flight mass spectrometry (PI-ReToF-MS), suggesting that the hitherto astronomically unobserved thiocarbonic acid represents a promising candidate for future astronomical searches. The formation of $H_2CS_3$ isomers were investigated through additional isotopically labeled experiments and the formation mechanisms through quantum chemical studies. These findings unravel a key reaction pathway to thiocarbonic acid and represent a first step toward its possible formation and detection in the ISM, shedding light on the missing sulfur problem.

## Introduction

Sulfur is the 10th most common element in the universe with an abundance of some $10^{-5}$ with respect to elemental hydrogen[1] and is one of the six principal elements of life besides hydrogen, carbon, nitrogen, oxygen, and phosphorus. As such, sulfur represents a key, fundamental element to the laboratory astrophysics and astrochemistry communities.[1–3] To date roughly 350, unique interstellar and circumstellar molecules have been identified including 47 sulfur-bearing molecules. Nearly a quarter of those (12) have been discovered within the last two years (Fig. 1).[4] In addition, the so-called sulfur depletion problem has raised many questions as to this element's fate in the interstellar medium (ISM). While diffuse clouds contain the same elemental sulfur proportions as its cosmic abundance,[3] the situation is quite distinct in dense regions. Carbon monosulfide (CS), sulfur monoxide (SO), and hydrogen disulfide ($H_2S$) have been detected in the gas phase in dense molecular clouds,[5] but these sulfur-bearing molecules account for only 1 % of the cosmic abundance of sulfur. The rest cannot be attributed to atomic sulfur.[6] This leads to the unanswered question of where 'missing' sulfur remains: the sulfur depletion problem. The same kind of irregularity has been reported in the solid phase — ices on interstellar nanoparticles (grains) in molecular clouds — as the only two molecules that have been conclusively detected under such circumstances are sulfur dioxide ($SO_2$)[7] and carbonyl sulfide (OCS);[8] these account for less than 5 % of the predicted sulfur abundance.[1] Other sulfur sinks almost certainly are waiting to be identified.

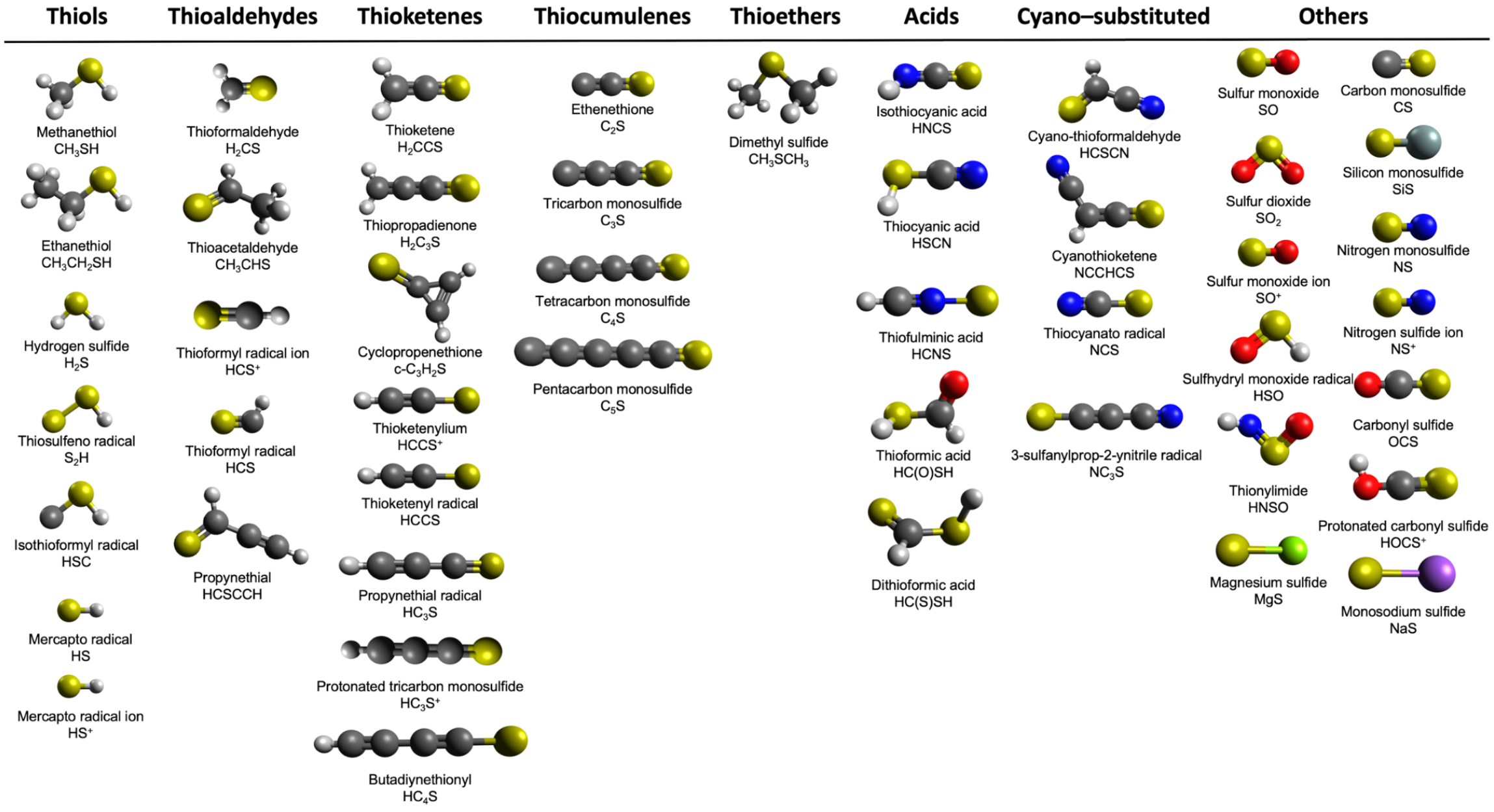

**Fig. 1** Key classes of sulfur-bearing molecules identified in the ISM. The colors correspond to the following elements: hydrogen (white), carbon (gray), nitrogen (blue), oxygen (red), sulfur (yellow), silicon (mint), magnesium (green) and sodium (purple).

Even though $H_2S$ is thought to be the main sulfur reservoir according to astrochemical models, it has not been detected in interstellar ices.[9] However, $H_2S$ has been identified in recent James Webb Space Telescope (JWST) surveys,[10] as well as toward protostars such as IRAS 16293-2422,[11] in Uranus' atmosphere,[12] and in comets such as IRAS-Araki-Alcock,[13] Austin,[14] and 67-P/Churyumov-Gerasimenko.[15] On the other hand, carbon disulfide ($CS_2$) has been detected toward several sources such as the sub-Neptune exoplanet TOI 270d,[16] comet 122P/de Vico,[17] and in Venus' atmosphere.[18] Additionally, previous studies have shown that OCS, which again is one of the two sulfur-bearing molecules detected in interstellar ices, can be formed within irradiated $CS_2$–$O_2$ ices.[19] Thus, the investigation of the sulfur chemistry in $CS_2$-containing interstellar ices represents a vital step for understanding the contributions that sulfur has in astrochemistry.

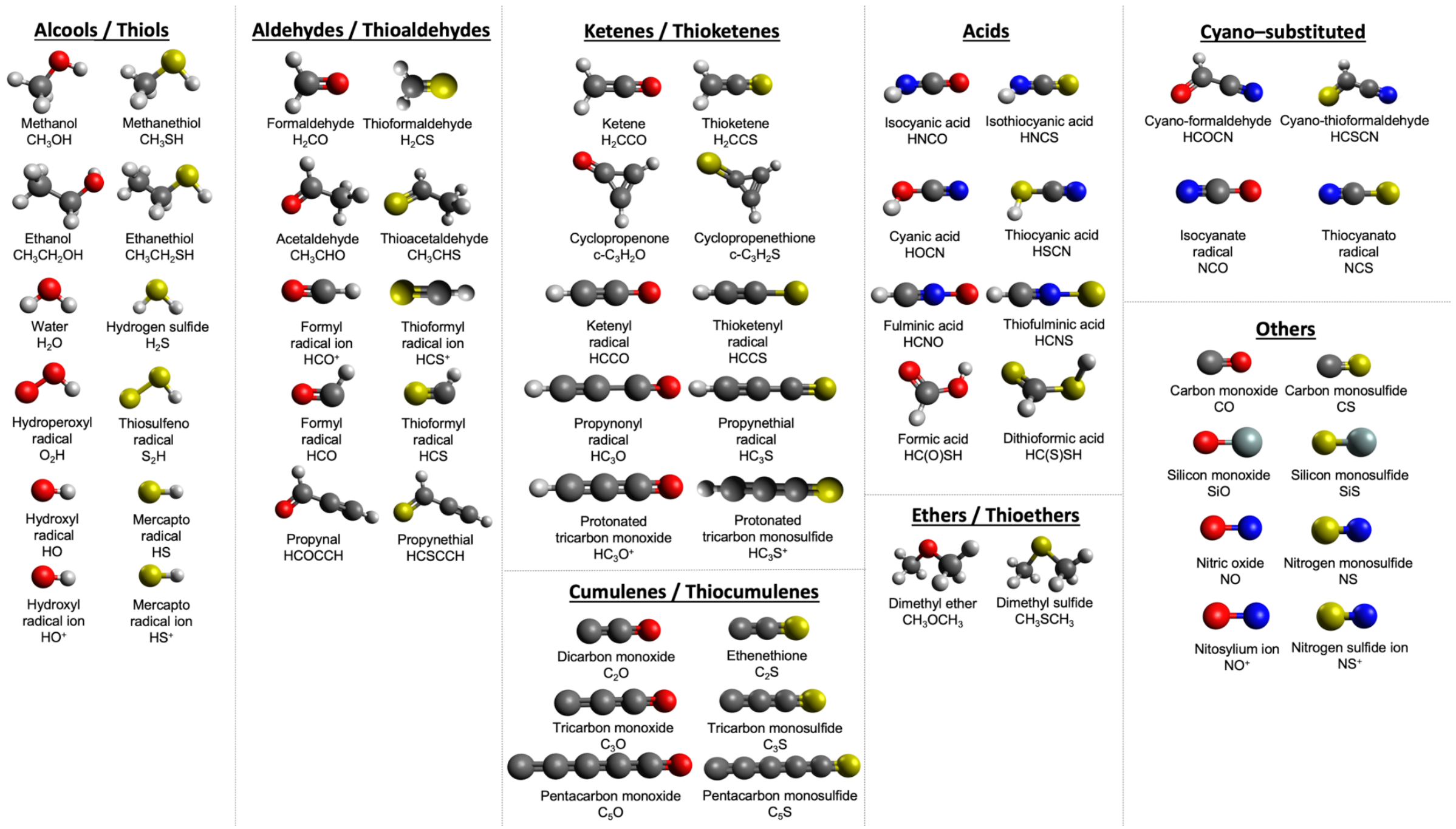

**Fig. 2** Key classes of oxygen/sulfur pair molecules identified in the ISM. The colors correspond to the following elements: hydrogen (white), carbon (gray), nitrogen (blue), oxygen (red), sulfur (yellow), and silicon (mint).

With the recent detection of carbonic acid ($H_2CO_3$) toward the Galactic center molecular cloud G+0.693–0.027[20] along with the known molecules discussed above containing C−S bonds, a sensible astrochemical target is the sulfur derivative of carbonic acid: thiocarbonic acid ($H_2CS_3$, **1**). Furthermore, the carbonic acid ($H_2CO_3$) - thiocarbonic acid ($H_2CS_3$) pair could represent the 31$^{st}$ pair, whose oxygen *and* sulfur analogs have been detected in the interstellar medium (Fig. 2). Sulfur counterparts of oxygenated molecules exhibit properties that are noticeably different stemming from the electronegativity difference between sulfur ($\chi$(S) =

2.64)[21] and oxygen ($\chi$(O) = 3.61),[21] and the fact that sulfur (R = 87 pm)[22] is significantly bigger than oxygen (R = 47 pm);[22] consequently, S–H bonds are less polarized than O–H ones, and are weaker hydrogen bond donors.[23] Even though properties can be different between those two atoms, their reactivity remains similar as they both belong to the chalcogen family. Indeed, a previous study by Wang *et al.* revealed that replacing CO–$H_2O$ precursors with CO–$H_2S$ yielded the sulfur counterpart of formic acid.[24] Similar to the proposed formation pathways of thioformic acid[24] and carbonic acid in interstellar ices analogs,[25,26] the interstellar formation pathway of thiocarbonic acid (**1**) may proceed within icy interstellar grains through the simple sulfur-bearing precursors $H_2S$ and $CS_2$. $H_2S$ can undergo homolytic cleavage under electron irradiation — proxies of galactic cosmic rays (GCRs)[27] — and lead to the formation of atomic hydrogen ($\dot{H}$) plus a mercapto ($\dot{S}H$) radical (Equation (1)); the suprathermal hydrogen atom can add to the C=S bond on either sulfur atom in $CS_2$ forming the mercaptothioxo radical (HS$\dot{C}$S) (Equation (2)); the barrier to addition of 6 kJ mol$^{-1}$ can be overcome by the excess kinetic energy of the suprathermal hydrogen atom;[28] and the thiocarbonic acid (**1**) could then form via the radical-radical recombination between the $\dot{S}H$ and HS$\dot{C}$S radicals (Equation (3)).

$$\mathrm{H_2S} \rightarrow \dot{\mathrm{H}} + \dot{\mathrm{S}}\mathrm{H} \quad (1) \quad \Delta_r G = +378\ \mathrm{kJ\ mol^{-1}}$$

$$\dot{\mathrm{H}} + \mathrm{CS_2} \rightarrow \mathrm{HS\dot{C}S} \quad (2) \quad \Delta_r G = -65\ \mathrm{kJ\ mol^{-1}}$$

$$\dot{\mathrm{S}}\mathrm{H} + \mathrm{HS\dot{C}S} \rightarrow \mathrm{H_2CS_3} \quad (3) \quad \Delta_r G = -280\ \mathrm{kJ\ mol^{-1}}$$

An alternative reaction pathway following $H_2S$ homolysis (Equation 1) is shown in Equations (2') and (3'). However, based on the study from Zheng and Kaiser on the formation of carbonic acid,[25] this pathway leading to the formation of thiocarbonic acid was ruled out, as in our case reaction (2') is 43 kJ mol$^{-1}$ less exoergic than reaction (2). Although Gattow and Krebs, and Jiaxuan have reported the transient formation of thiocarbonic acid (**1**) in aqueous solution from $BaCS_3$/HCl[29] as well as from liquid $CS_2$ in a mixture of solvents,[30] laboratory experiments have thus far failed to detect thiocarbonic acid (**1**) in ice analogs under astrochemical conditions.

$$\dot{\mathrm{S}}\mathrm{H} + \mathrm{CS_2} \rightarrow \mathrm{HSC\dot{S}} \quad (2') \quad \Delta_r G = -22\ \mathrm{kJ\ mol^{-1}}$$

$$\dot{\mathrm{H}} + \mathrm{HSC\dot{S}} \rightarrow \mathrm{H_2CS_3} \quad (3') \quad \Delta_r G = -323\ \mathrm{kJ\ mol^{-1}}$$

Here we present the first preparation of thiocarbonic acid (**1**) in interstellar ice analogs composed of hydrogen sulfide and carbon disulfide. Electron irradiation was applied to the ice mixtures at a low temperature of 5 K, simulating the effect of GCRs on icy grains in a cold molecular cloud over some $10^6$ years.[27] While heated from 5 to 330 K during the temperature program desorption (TPD) phase, thiocarbonic acid (**1**) was detected in the gas phase through *isomer-specific* vacuum ultraviolet (VUV) photoionization coupled with reflectron time-of-flight mass spectrometry (PI-ReToF-MS).[31] These results suggest that thiocarbonic acid (**1**) represents a suitable candidate for gas-phase searches through radio telescopes such as the Atacama Large Millimeter/submillimeter Array (ALMA). Additionally, our findings reveal a key formation pathway of thiocarbonic acid (**1**) thus enhancing our fundamental understanding of the sulfur chemistry within ices of cold molecular clouds. These findings not only help us better understand the sulfur depletion problem, but also provide insights into the interstellar formation of complex organic chemistry via non-equilibrium chemistry, especially the synthesis of sulfur-bearing molecules in deep space.

## Methods

### *Experimental Section*

The experiments were conducted in an ultra-high vacuum (UHV) stainless steel main chamber at a pressure of a few $10^{-11}$ Torr obtained by turbomolecular pumps (Osaka, TG1300MUCWB and TG420MCAB), which are backed by a dry scroll pump (XDS35i, BOC Edwards).[32] The chemical samples used for the experiments were hydrogen sulfide ($H_2S$, Sigma-Aldrich, >99.5%), hydrogen sulfide-d$_2$ ($D_2S$, Sigma-Aldrich, 97 atom % D), liquid carbon disulfide ($CS_2$, Sigma-Aldrich, > 99.9%), and carbon disulfide-$^{13}$C ($^{13}CS_2$, Sigma-Aldrich, 99 atom % $^{13}$C). The carbon disulfide sample was stored in a borosilicate vial and underwent several freeze-pump-thaw cycles to remove residual atmospheric gases. The hydrogen sulfide and carbon disulfide were premixed in a separate UHV chamber at a ratio of (3.8 ± 0.5):1 for $H_2S$:$CS_2$. A polished silver substrate was mounted on an oxygen-free copper cold finger and was cooled down to 5 K using a closed-cycle helium refrigerator (Sumitomo Heavy Industries, RDK-415E).[24] The premixed gas was then introduced into the main chamber at a pressure of $5 \times 10^{-8}$ Torr via a glass capillary array and deposited onto the substrate that can be rotated and translated vertically. During deposition, the ice growth was monitored *in situ* via laser interferometry with a helium-neon laser (Melles Griot; 25-LHP-230) operating at 632.8 nm.[33] The ice thickness was calculated based on the refractive index (n) of amorphous hydrogen sulfide (n = 1.41 ± 0.02) at 19 K[34] and that of carbon disulfide (n = 1.78 ± 0.02) at 10 K. The latter was determined using two methods: the Lorentz-Lorentz equation (n = 1.78 ± 0.02)[35] and the thermo-optic coefficient (n = 1.77 ± 0.02).[36] According to the concentration-weighted average refractive index (n = 1.49 ± 0.02) and the ice composition, the deposition ice thickness was determined to be 850 ± 50 nm (Table S1). Based on the densities of hydrogen sulfide (1.1 g cm$^{-3}$)[37,38] and carbon disulfide (1.54 g cm$^{-3}$),[39] the average penetration depth of electrons in $H_2S-CS_2$ ice was calculated to be 290 ± 50 nm using Monte Carlo simulations with CASINO v2.48 software suite.[40] The ice thickness (850 ± 50 nm) is significantly greater than the average penetration depth, effectively preventing electron-initiated interactions between the ice and the silver substrate. The relative concentrations of reactants in the ices were measured using a Fourier transform infrared (FTIR) spectrometer (Thermo Nicolet 6700, 4 $cm^{-1}$ resolution) in the range of 6000−500 $cm^{-1}$. Specifically, the integrated infrared absorption bands of hydrogen sulfide at 2546 $cm^{-1}$ ($\nu_1$, $1.12 \times 10^{-17}$ cm $molecule^{-1}$)[38] and carbon disulfide at 1499 $cm^{-1}$ ($\nu_3$, $9.13 \times 10^{-17}$ cm $molecule^{-1}$)[19] were used.

After deposition, the ice mixtures were subjected to electron irradiation (SPECS, EQ PU-22, 5 keV) at a current of 20 ± 1 nA for 10 minutes over an area of 160 $mm^2$ at an incidence angle of 70°. The irradiation conditions correspond to doses up to 0.25 eV $molecule^{-1}$ for hydrogen sulfide and up to 1.0 eV $molecule^{-1}$ for carbon disulfide, simulating secondary electrons generated along the tracks of GCRs in cold molecular clouds aged $(1.0 \pm 0.3) \times 10^6$ years.[27] To monitor changes in chemical composition, FTIR spectra were recorded before, during, and after the irradiation. Post-irradiation analysis of the subliming products from the ice mixtures was carried out using single photon vacuum ultraviolet photoionization coupled with reflectron time-of-flight mass spectrometry (PI-ReToF-MS). This technique enabled the isomer selective, sensitive detection of the chemical species released during the temperature-programmed desorption (TPD) phase. The irradiated ice samples were gradually heated from 5 K to 330 K at a rate of 1 K $min^{-1}$. The subliming species were photoionized 2 mm above the ice surface using a pulsed VUV light source (30 Hz). The VUV photons at 10.49, 9.34, 8.90, 8.81, 8.66, and 8.17 eV were generated through four-wave mixing (FWM) using two synchronized pulsed laser beams, which were produced by two dye lasers (Sirah, Cobra-Stretch) pumped by two Nd:YAG (neodymium-doped yttrium aluminum garnet) lasers (Spectra-Physics, Quanta Ray Pro 250−30 and 270−30, 30 Hz).[41] The 10.49 eV photons were produced using the triple third harmonic ($\omega_{VUV} = 3\omega_1$; $\omega_1$, 355 nm) of the Nd:YAG laser. The remaining photon energies were produced by difference FWM ($\omega_{VUV} = 2\omega_1 - \omega_2$) in pulsed jets of xenon or krypton gas as a nonlinear medium (Table S2). The VUV photons were separated from lower energy fundamentals using a biconvex lithium fluoride lens (Korth Kristalle, R1 = R2 = 131 mm) in an off-axis geometry. Each recorded TPD profile was corrected for fluctuations in the VUV flux. The resulting ions were extracted within a reflectron time-of-flight mass spectrometer (Jordan TOF Products, Inc.) and detected with a dual microchannel plate (MCP) detector in the chevron configuration. Ion signals were amplified with a preamplifier (Ortec, 9305) and recorded by a multichannel scaler (FAST ComTec, MCS6A). For each mass spectrum, the accumulation time of ion signals was 2 min (3600 sweeps) with an ion arrival time accuracy of 3.2 ns.[32] Additional blank experiment with unirradiated $H_2S$−$CS_2$ ice was performed at 10.49 eV.

### *Computational Section*

Adiabatic ionization energies and relative stabilities of distinct $H_2CS_3$ isomers were computed exploiting the MOLPRO 2022.1 quantum chemistry software[42–44] and coupled cluster singles and doubles with perturbative triple excitations with the F12 correction (CCSD(T)-F12)[45] conjoined to the correlation-consistent polarized valence triple-zeta basis set tailored for F12 methods (cc-pVTZ-F12).[46] For the dimers and transition states, the geometries were optimized within the Gaussian 16 software[47] with density functional theory (DFT) via ωB97XD[48] alongside the augmented correlation-consistent polarized valence triple-zeta basis set (aug-cc-pVTZ).[46] From these geometries, CCSD(T)-F12b/cc-pVTZ-F12[45,46] single point energies were computed within the MOLPRO 2022.1 quantum chemistry software.[42–44] The method used to optimize the geometry also provides the zero-point vibrational energy (ZPVE) corrections which are added to the coupled cluster energies. Adiabatic ionization energies (IEs) are determined by differences between the respective energies of the optimized neutral isomers and the cations. The IEs employ error analysis which takes into account the computational uncertainty of ± 0.04 eV[49] and a correction of −0.03 eV caused by the electric-field-induced Stark effect.[24] The same adiabatic approach was employed for computations of the relative energies for the related isomers/conformers.

## Results and Discussion

### *Infrared Spectroscopy*

Fourier-transform Infrared spectroscopy (FTIR) is utilized to observe the emergence of functional groups of complex organic molecules during the radiation exposure of the hydrogen sulfide ($H_2S$) and carbon disulfide ($CS_2$) ices. Figure 3 displays the FTIR spectra before (black) and after (red) the irradiation of the $H_2S$−$CS_2$ ices along with their respective assignments. Prior to the irradiation, all absorption bands in the spectrum can be attributed to hydrogen sulfide and carbon disulfide, such as the S−H stretching modes ($\nu_s$ ($\nu_1$ and $\nu_3$), 2546 cm$^{-1}$) and bending mode ($\nu_2$, 1166 cm$^{-1}$) of hydrogen sulfide[50,51] as well as the SCS antisymmetric stretching mode ($\nu_3$, 1499 cm$^{-1}$) of carbon disulfide.[52–54] Key features are also associated with combination bands. For example, the band at 2153 cm$^{-1}$, present prior to the irradiation, can be attributed to a combination of carbon disulfide modes, as noted by previous studies. Plyer reports that the $\nu_1$ + $\nu_3$ combination band should appear around 2160 cm$^{-1}$ in gas-phase $CS_2$,[52] which is close to the 2153 cm$^{-1}$ observed in the $H_2S$−$CS_2$ spectrum in the ices. Following irradiation, only a small, new absorption feature was observed in the 2530−2200 cm$^{-1}$ region with a slight increase in band intensity (Fig. 3).

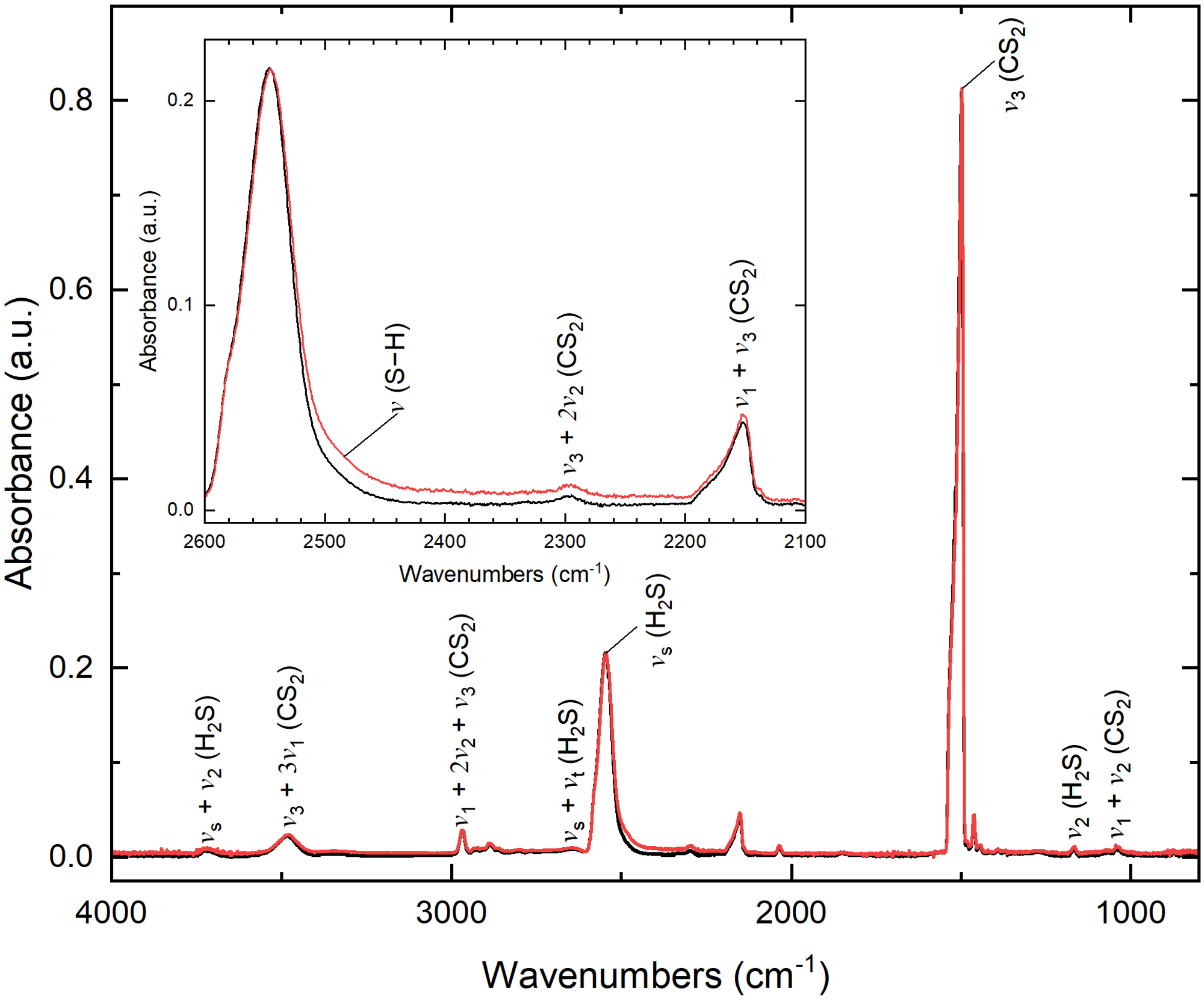

**Fig. 3** IR spectrum of pristine (black) and irradiated (red) $H_2S$−$CS_2$ ice at 5 K. Detailed assignments are listed in Table S3. Inserted figure shows the zoomed region from 2600 to 2100 cm$^{-1}$ revealing a broader signal after irradiation.

FTIR spectra before (black) and after (red) the irradiation of isotopically-labeled ices ($H_2S$–$^{13}CS_2$ and $D_2S$–$CS_2$) were also performed. Prior to the irradiation of the isotopic $H_2S$−$^{13}CS_2$ ice, all absorption bands in the spectrum can be attributed to hydrogen sulfide and isotopic carbon disulfide (Fig. S1). Similarly to the non-isotopic ice, S−H stretching modes ($\nu_s$ ($\nu_1$ and $\nu_3$), 2546 ± 5 cm$^{-1}$) and bending mode ($\nu_2$, 1166 ± 5 cm$^{-1}$) of hydrogen sulfide are present.[50,51] Regarding isotopic carbon disulfide, the strong SCS antisymmetric stretching vibration ($\nu_3$, 1460 cm$^{-1}$) has shifted of 39 cm$^{-1}$ toward low frequency, which is coherent with $^{13}CS_2$ spectrum reported in previous literature.[54] Prior to the irradiation of the isotopic $D_2S$−$CS_2$ ice, all absorption bands in the spectrum can be attributed to deuterium sulfide, residual hydrogen sulfide, and carbon disulfide (Fig. S1). The SCS antisymmetric stretching mode appears at the same frequency ($\nu_3$, 1499 ± 5 cm$^{-1}$) than the non-isotopic ice, and new peaks corresponding to deuterium sulfide are observed. For example, the S−H / S−D stretching modes ($\nu_s$ ($\nu_1$ and $\nu_3$)) shift from 2546 cm$^{-1}$ for residual hydrogen sulfide to 1847 cm$^{-1}$ for deuterium sulfide which is coherent to what has been reported previously.[51] Following the irradiation, only a slight increase in intensity in the 2530−2200 cm$^{-1}$ region for the $H_2S$−$^{13}CS_2$ is observed. Table S3 collects the IR absorption features of the pristine and irradiated $H_2S$−$CS_2$, $H_2S$–$^{13}CS_2$ and $D_2S$–$CS_2$ ices. Given that experiments are conducted at a low-dose irradiation (20 nA, 10 min) and absorption features of the formed organics overlap, FTIR spectroscopy alone cannot uniquely detect complex compounds such as thiocarbonic acid (**1**), a more sensitive and *isomer-specific* technique is necessary to investigate the reaction products and the mechanism pathway.

**Table 1** Adiabatic ionization energies (IEs) and relative energies (Rel. E) of $H_2CS_3$ isomers ($m/z$ = 110) were computed at the CCSD(T)-F12/cc-pVTZ-F12 level of theory. The IE ranges are corrected for the combined error limits of ± 0.04 eV and the thermal and Stark effect by −0.03 eV.

| Isomer | | Structure | | Rel. E (kJ/mol) | IE (eV) | IE range with error (eV) | IE range corrected with Stark effect (eV) |
|---|---|---|---|---|---|---|---|
| **1** | HSC(=S)SH | ct: **1a** | 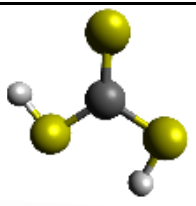 | 1 | 8.88 | 8.84–8.92 | 8.81–8.89 |
| | | cc: **1b** | 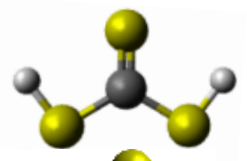 | 0 | 8.87 | 8.83–8.91 | 8.80–8.88 |
| | | tt : **1c** | 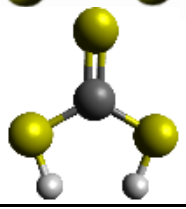 | 8 | 8.78 | 8.74–8.82 | 8.71–8.79 |
| **2** | HSSC(=S)H | | 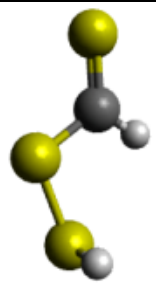 | 12 | 8.97 | 8.93–9.01 | 8.90–8.98 |
| **3** | HSCHSS | | 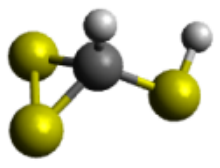 | 38 | 8.52 | 8.46–8.54 | 8.43–8.51 |
| **4** | $H_2CS_3$ (ring) | | 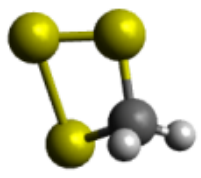 | 51 | 8.00 | 7.96–8.04 | 7.93–8.01 |

### *Mass Spectrometry*

The PI-ReToF-MS approach was utilized to identify the subliming products of the irradiated hydrogen sulfide−carbon disulfide ices. PI-ReToF-MS represents an isomer-selective detection technique based on the desorption temperature of the molecules and their ability to ionize, which occurs only when the photon energy is higher than the adiabatic ionization energy (IE) of products.[31] This requires the knowledge of the adiabatic ionization energies (IEs) of isomers of $H_2CS_3$ (Table 1). These data account for distinct isomers and conformers; they are also corrected for the thermal and Stark effect, which can reduce effective IEs by up to 0.03 eV. These isomers can be distinguished in separate experiments exploiting tunable VUV photons. Six VUV photon energies at 10.49, 9.34, 8.90, 8.81, 8.66, and 8.17 eV were selected to distinguish the four isomers (**1**, **2**, **3**, and **4**) (Fig. 4). Photons at 10.49 and 9.34 eV are capable of ionizing all isomers if present. At 8.90 eV, only isomers **1** (IE = 8.71−8.89 eV), **3** (IE = 8.43−8.51 eV) and **4** (IE = 7.93−8.01 eV) are ionizable. A photon of a reduced energy of 8.66 eV was chosen to ionize isomers **3** and **4** if formed. However, neither thiocarbonic acid (**1**) nor isomer **2** would be detected in the experiments with this photon energy. Likewise, 8.17 eV would only ionize isomer **4**.

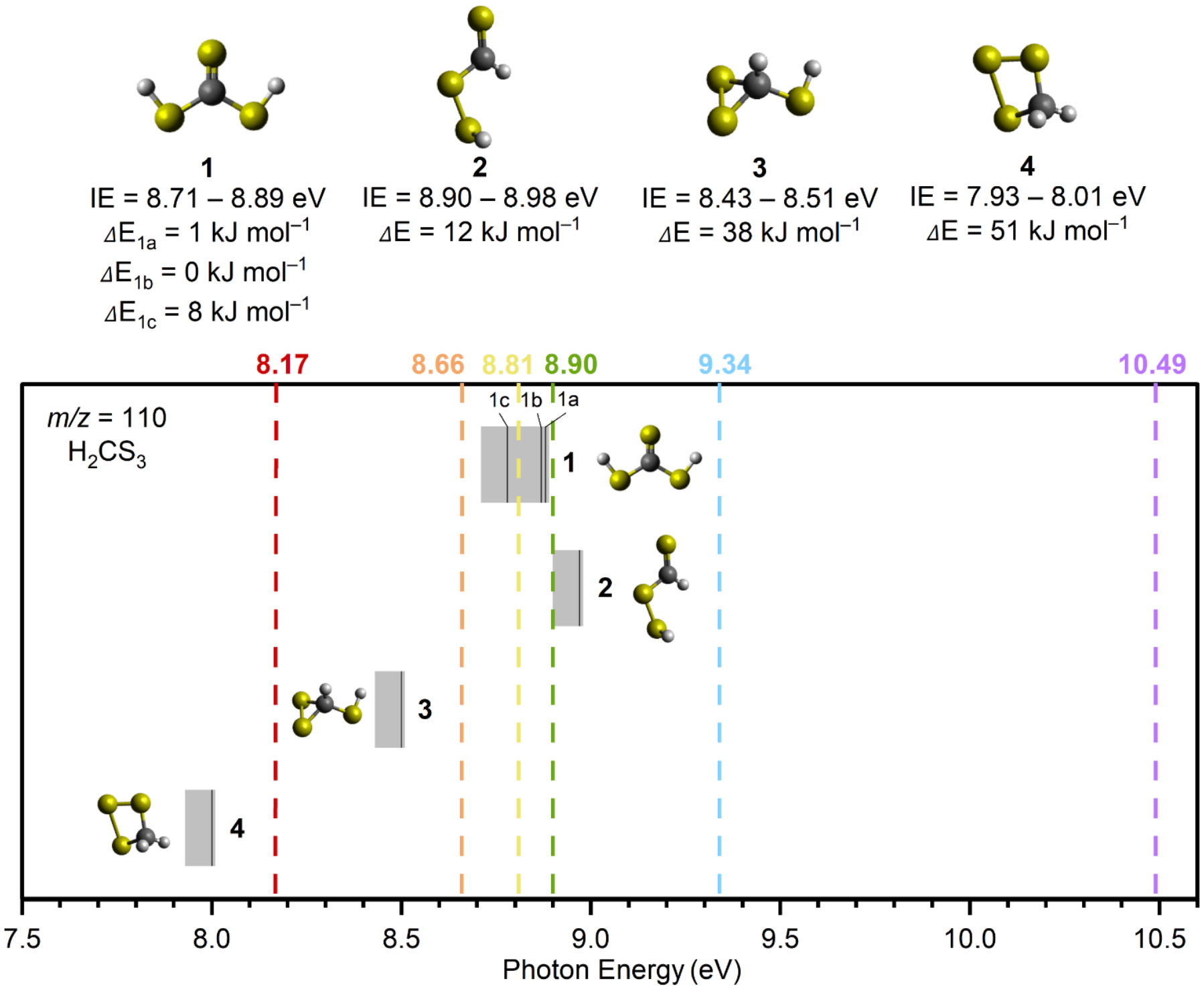

**Fig. 4** Computed ionization energies (IEs) of $H_2CS_3$ isomers (solid black lines) along with error limits (Table 1). Isomers **1** to **4** are represented on top of the figure along with their IEs and relative energies (ΔE). For isomer **1**, the three relative energies correspond to its conformers. Vacuum ultraviolet (VUV) photon energies (dashed lines) at 10.49, 9.34, 8.90, 8.81, 8.66, and 8.17 eV were used to selectively ionize the isomers **1** to **4** in the gas phase during TPD.

The PI-ReToF-MS data from the irradiated hydrogen sulfide−carbon disulfide ices are compiled in Figure 5 and analyzed to extract the TPD profiles of the mass of interest *m/z* = 110 for $H_2CS_3^+$ (Fig. 6). The TPD profile of *m/z* = 110 at 10.49 eV reveals sublimation events at 130 K (Peak I), 219 K (Peak II), and 285 K (Peak III) (Fig. 6a). A blank experiment was conducted without exposing the ices to electron irradiation; the remaining parameters remained unchanged (Fig. 5a). The signal at *m/z* = 110 for the blank experiment reveals a sublimation event at 130 K, but no sublimation event was detected at 219 K nor 285 K, indicating that Peak I results from the saturation of the detector upon $CS_2$ sublimation (Fig. S2) and Peaks II and III are caused by the electron irradiation of the ice mixtures.

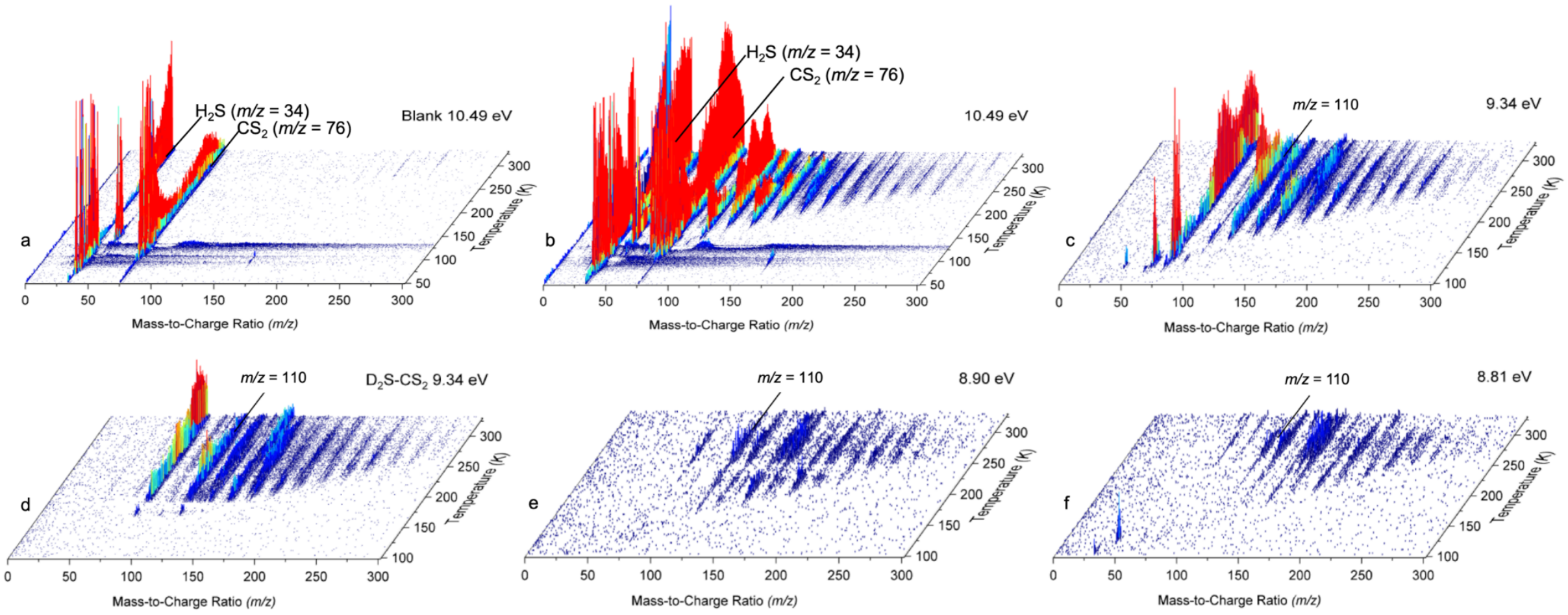

**Fig. 5** PI-ReToF-MS spectra collected during the TPD of hydrogen sulfide-carbon disulfide ices: unirradiated (blank) $H_2S-CS_2$ ice recorded at 10.49 eV (a), the irradiated $H_2S-CS_2$ ice recorded at 10.49 (b), 9.34 (c), 8.90 (e), and 8.81 eV (f), and the irradiated $D_2S-CS_2$ ice recorded at 9.34 eV (d).

In the irradiated ices, the signal at *m/z* = 110 can be assigned to molecules with the molecular formulae $C_9H_2$, $C_6H_6S$, $C_8H_{14}$ and/or $H_2CS_3$. Since the $^{34}S$ isotope contributes about 4.2 % of the sulfur isotopic composition, the presence of signal in the 200−240 K temperature range in the TPD profile at *m/z* = 112 for $H_2S$–$CS_2$ ice at 9.34 eV (Fig. S3) reveals the presence of at least one sulfur atom in the molecular formulae. Isotopically labeled reactants were exploited to fully assign the molecular formula. The experiment performed at 9.34 eV using fully deuterated ices $D_2S-CS_2$ shows a shift by 2 amu and hence a TPD profile at *m/z* = 112; this TPD matches the TPD profile of *m/z* = 110 for $H_2S-CS_2$ ice (Fig. 7) in the 200−240 K temperature range (Peak II). This finding indicates the presence of exactly two hydrogen atoms in the produced molecule. Another experiment at 9.34 eV using carbon-13 isotopically labeled ice ($H_2S-^{13}CS_2$) shows an isotopic mass shift by 1 amu in the TPD profile at *m/z* = 111 yielding a similar desorption pattern for Peak II; this finding accounts for the presence of only one carbon atom (Fig. 7). These results confirm the assignment of the sublimation event peaking at 219 K (Peak II) to the molecular formula $H_2CS_3$.

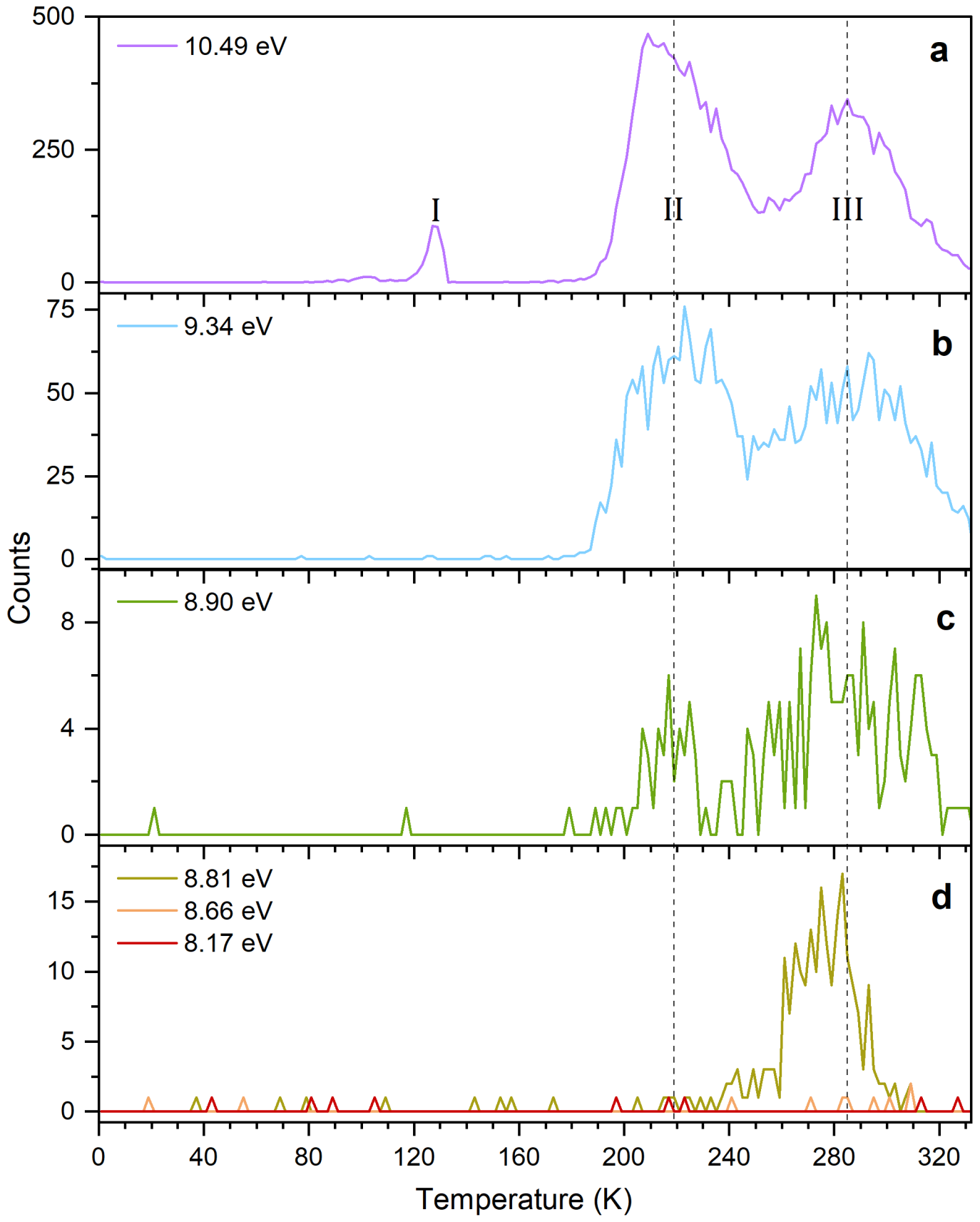

**Fig. 6** TPD profiles of $m/z$ = 110 from the irradiated $H_2S$−$CS_2$ ice at 10.49 (a), 9.34 (b), 8.90 (c), 8.81, 8.66, and 8.17 eV (d). The dashed black lines indicate the sublimation peaks II (219 K) and III (285 K).

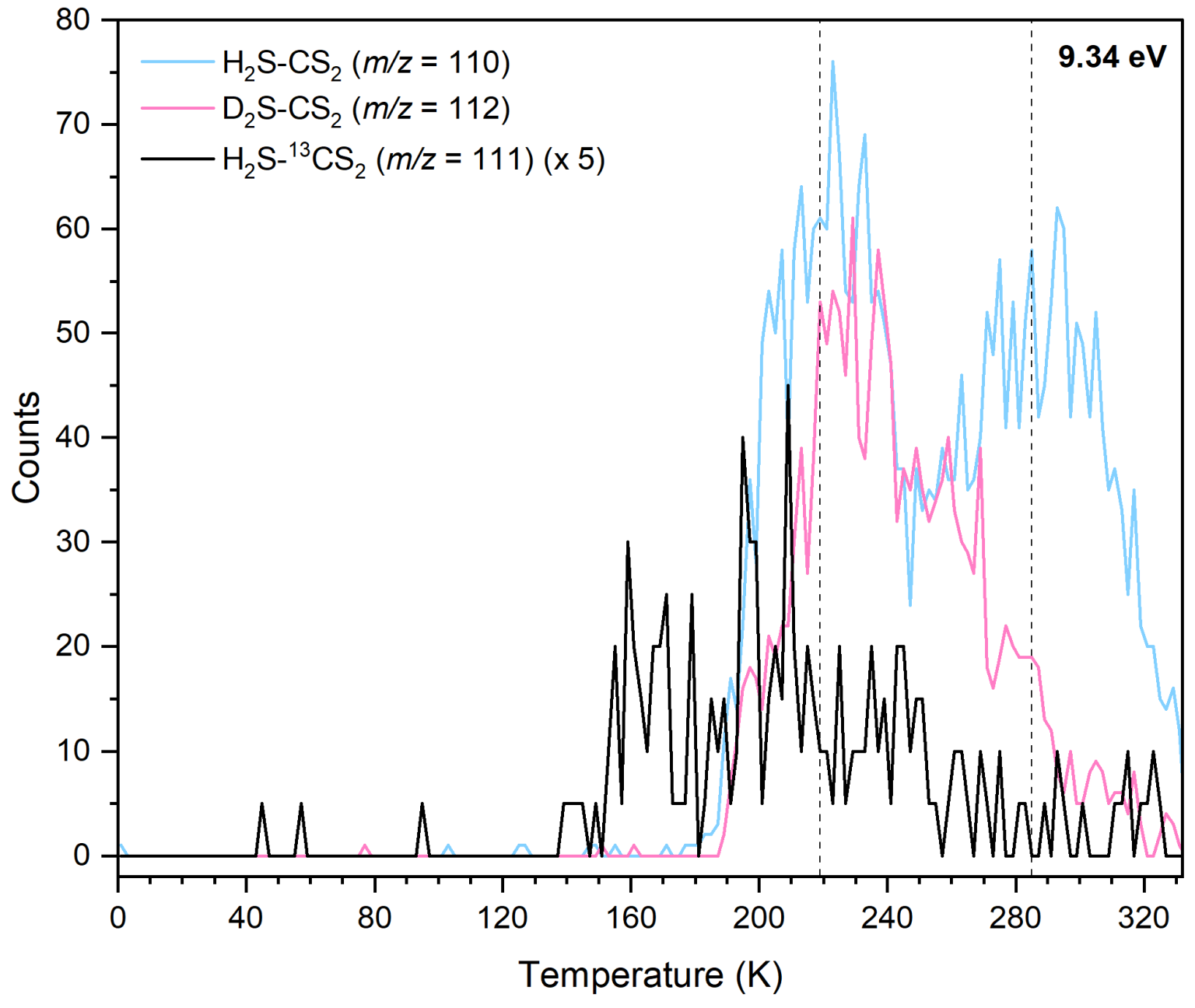

**Fig. 7** TPD profiles at 9.34 eV for isotopically labeled hydrogen sulfide-carbon disulfide ice mixtures: $H_2S$−$CS_2$ ($m/z$ = 110), $D_2S$−$CS_2$ ($m/z$ = 112), $H_2S$−$^{13}CS_2$ ($m/z$ = 111). The dashed lines indicate sublimation peaks II (219 K) and III (285 K).

Having identified $H_2CS_3$ isomer(s) as a carrier of the sublimation event at *m/z* = 110 for $H_2S-CS_2$ ices, we are elucidating now the nature of the isomer by tuning the VUV energy (Fig. 4). At 8.90 eV, the TPD profile at *m/z* = 110 ($H_2CS_3^+$) still shows the sublimation event at 219 K (Fig. 6c) indicating Peak II is not associated with isomer **2**. The photon energy was further reduced to 8.66 eV, for which no ion signal was detected at *m/z* = 110 (Fig. 6d), showing the absence of isomers **3** and **4** formation. Therefore, this selective identification reveals that Peak II at *m/z* = 110 ($H_2CS_3^+$) corresponds solely to thiocarbonic acid (**1**). Note that isomer **1** is the most stable isomer among all $H_2CS_3$ isomers.

**Table 2** Dissociative ionization energies (dissociation IEs) of thiocarbonic acid dimers ($H_4C_2S_6$) (top) and their adiabatic ionization energies (IEs) and relative energies (Rel. E) (bottom) were computed at the CCSD(T)-F12/cc-pVTZ-F12 + ZPVE(ωB97XD/aug-cc-pVTZ) level of theory. The calculated energies were corrected for the combined error limits of ± 0.04 eV and the thermal and Stark effect by −0.03 eV.

Dissociative ionization energies of thiocarbonic acid dimers leading to isomer **1** ($H_2CS_3$) cations: *m/z* = 110, $H_2CS_3^+$

| | Conformer | Structure | Dissociation IE (eV) | IE range with error (eV) | Corrected IE with Stark effect (eV) |
|---|---|---|---|---|---|
| **1a**$^+$ | ct | 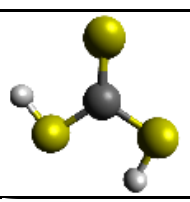 | 9.10 | 9.06–9.14 | 9.03–9.11 |
| **1b**$^+$ | cc | 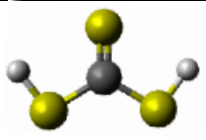 | 9.08 | 9.04–9.12 | 9.01–9.09 |
| **1c**$^+$ | tt | 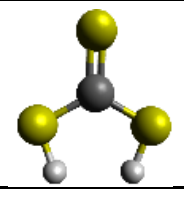 | 9.02 | 8.98–9.06 | 8.95–9.03 |

IEs of three dimers of isomer **1**: *m/z* = 220, $H_4C_2S_6$

| | Conformer | Structure | Rel. E (kJ/mol) | IE (eV) | IE range with error (eV) | Corrected IE with Stark effect (eV) |
|---|---|---|---|---|---|---|
| **1a-1a** | ct-ct | 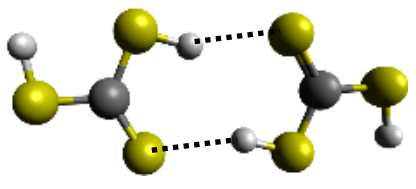 | 1 | 7.93 | 7.89–7.97 | 7.86–7.94 |
| **1b-1b** | cc-cc | 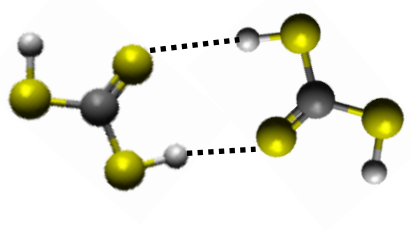 | 0 | 8.93 | 8.89–8.97 | 8.86–8.94 |
| **1c-1c** | tt-tt | 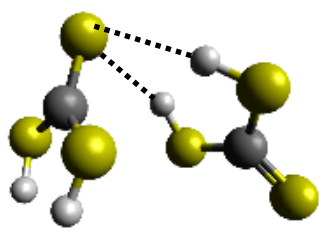 | 14 | 8.32 | 8.28–8.36 | 8.25–8.33 |

Finally, we are exploring the nature of sublimation event III; this might be caused by the sublimation of dimers of thiocarbonic acid (220 amu); their dissociative photoionization could result in signal at *m/z* = 110.[55] IEs were calculated for dimer structures focusing on two of the three most stable ones (**1a-1a** and **1b-1b** as dimers **1a-1a** and **1b-1a** have the same IEs (Table S4)) and the least stable dimer structure we found (**1c-1c**), in order to investigate dimer formation and detection with regard to their stability.[56] The results are compiled in Table 2 and Figure 8: **1a-1a** (IE = 7.86−7.94 eV; ΔE = 1 kJ $mol^{-1}$), **1b-1b** (IE = 8.86−8.94 eV; ΔE = 0 kJ $mol^{-1}$), and **1c-1c** (IE = 8.25−8.33 eV; ΔE = 14 kJ $mol^{-1}$). At *m/z* = 220, signal is collected at photon energies of 10.49, 9.34, 8.90, 8.81, or 8.66 eV for the sublimation event in the temperature range of 240−300 K; this range matches Peak III at *m/z* = 110 (Fig. 9 and Fig. S4). However, at 8.17 eV, where only dimer **1a-1a** can be ionized if present, no sublimation event is detected in the TPD profile of *m/z* = 220 (Fig. S4d) revealing dimer **1a-1a** is not formed. At 8.66 and 8.81 eV, dimer **1c-1c** (IE = 8.25−8.33 eV) can be ionized if present. The TPD signal of *m/z* = 220 for these two VUV photon energies shows a weak sublimation event (Fig. S4d). These findings indicate that the sublimation event at *m/z* = 220 from 240 to 300 K corresponds to the sublimation of $H_2CS_3$ dimer **1c-1c**. Considering that this sublimation event happens at similar temperature as Peak III, the latter might result from dissociative photoionization of the sublimed dimers. Although a sublimation event at 285 K in the TPD profile of *m/z* = 110 is detected during experiments at 8.90 eV and 8.81 eV (Fig. 6), the calculated dissociative IEs of the corresponding dimers exceed 8.90 eV (Table 2 and Fig. 8). Therefore, Peak III accounts for the dissociative photoionization of the sublimed dimer **1c-1c** and/or other possible dimer structures or even larger oligomers in which one monomer can dissociate more easily.

Recent literature reported that carbamic acid dimers sublimate 35 K higher than the monomers.[57] This trend is also supported in our system where sublimation of thiocarbonic acid dimers commences 66 K higher than the monomers. The additional temperature difference can be rationalized in terms of a polymer layer generated during the radiation exposure of the exposed ices. This in turn increases the sublimation temperature.

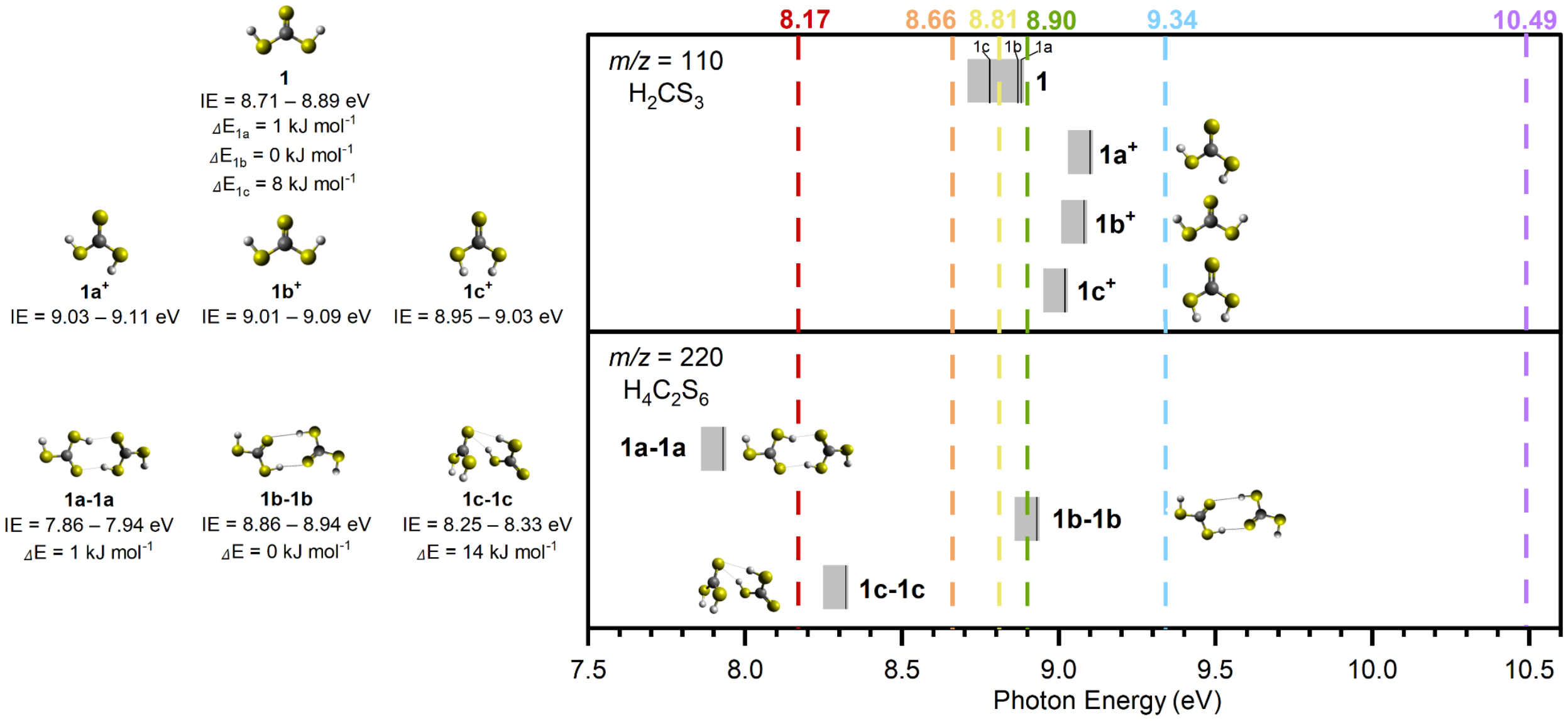

**Fig. 8** Computed ionization energies (IEs; solid black lines) of thiocarbonic acid (1; H2CS3) conformers (top) and their corresponding dimers (bottom) along with error limits (Table 2). The conformers marked with an asterisk are formed upon dissociative sublimation of their corresponding dimers. Conformers of thiocarbonic acid and their corresponding dimers are represented on the left, along with their relative energies and ionization energies. VUV photon energies (dashed lines) at 10.49, 9.34, 8.90, 8.81, 8.66, and 8.17 eV were used to selectively ionize the subliming molecules during the TPD process.

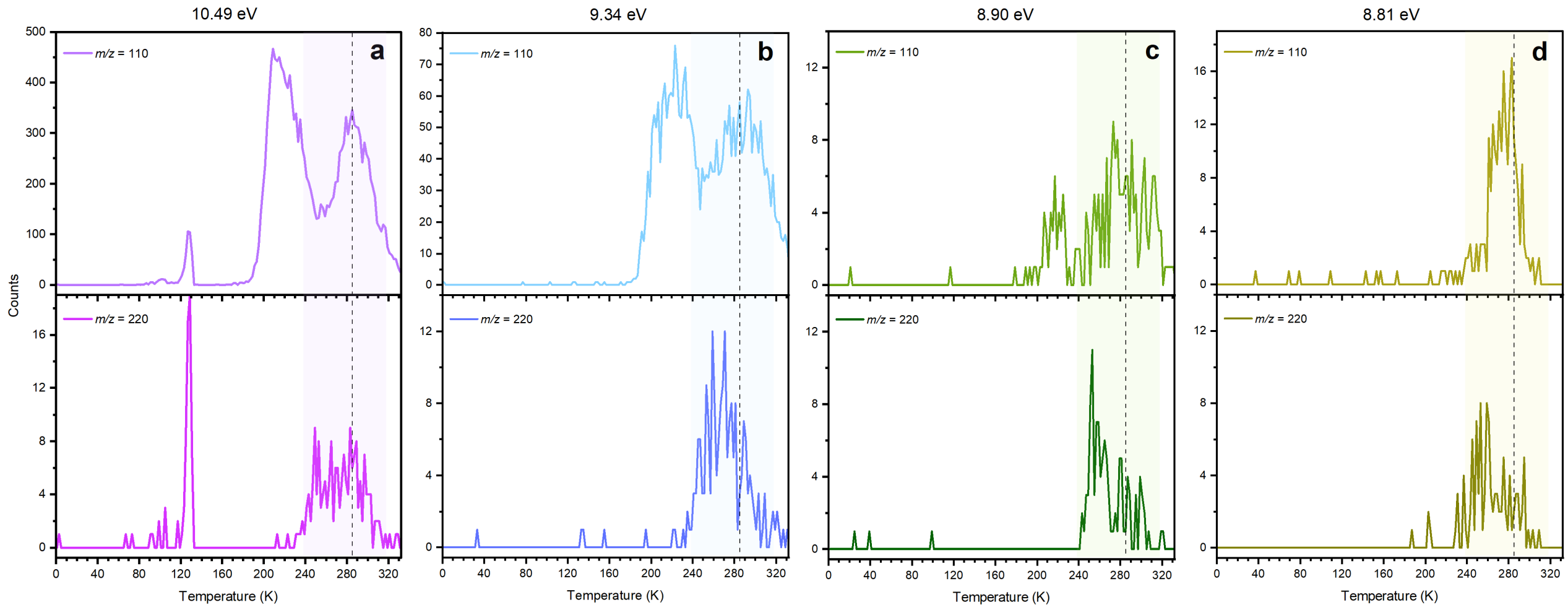

**Fig. 9** Compared TPD profiles of *m/z* = 110 and *m/z* = 220 from the irradiated $H_2S-CS_2$ ice recorded at 10.49 (a), 9.34 (b), 8.90 (c), and 8.81 eV (d). The dashed line indicates the sublimation peak at 285K (Peak III).

## *Discussion*

Having provided compelling evidence for the formation of thiocarbonic acid (**1**) under astrochemical conditions, we now focus on its potential formation pathway based on gas phase computational studies. The results were applied to our condensed phase experiment keeping in mind that energies might slightly differ due to molecular interactions. However, sulfur being a weaker hydrogen bond donor compared to oxygen,[23] sulfur-containing ices are less organized in the condensed phase and energy values may be consequently closer to gas phase calculations. Similar to the formation of carbonic acid ($H_2CO_3$) from processed $H_2O$−$CO_2$ ice analogs,[25] the formation pathway for thiocarbonic acid (**1**) in irradiated hydrogen sulfide−carbon disulfide ices is proposed in Scheme 1. First, hydrogen sulfide is radiolyzed to a mercapto radical (ṠH) by the loss of a hydrogen atom (Ḣ) through an endoergic reaction requiring 378 kJ mol$^{-1}$.[24] This energy can be supplied with the energetic electrons during irradiation. Then, the addition of Ḣ onto carbon disulfide at the sulfur site leads to the formation of a mercaptothioxo radical (HSĊS) (radical 1 in Fig. 10A) after passing an entrance barrier of 6 kJ mol$^{-1}$ calculated at the CCSD(T)-F12/cc-pVTZ-F12 level of theory. The reaction leading to radical 1 is exoergic by 65 kJ mol$^{-1}$ (addition [1], Fig. 10A).

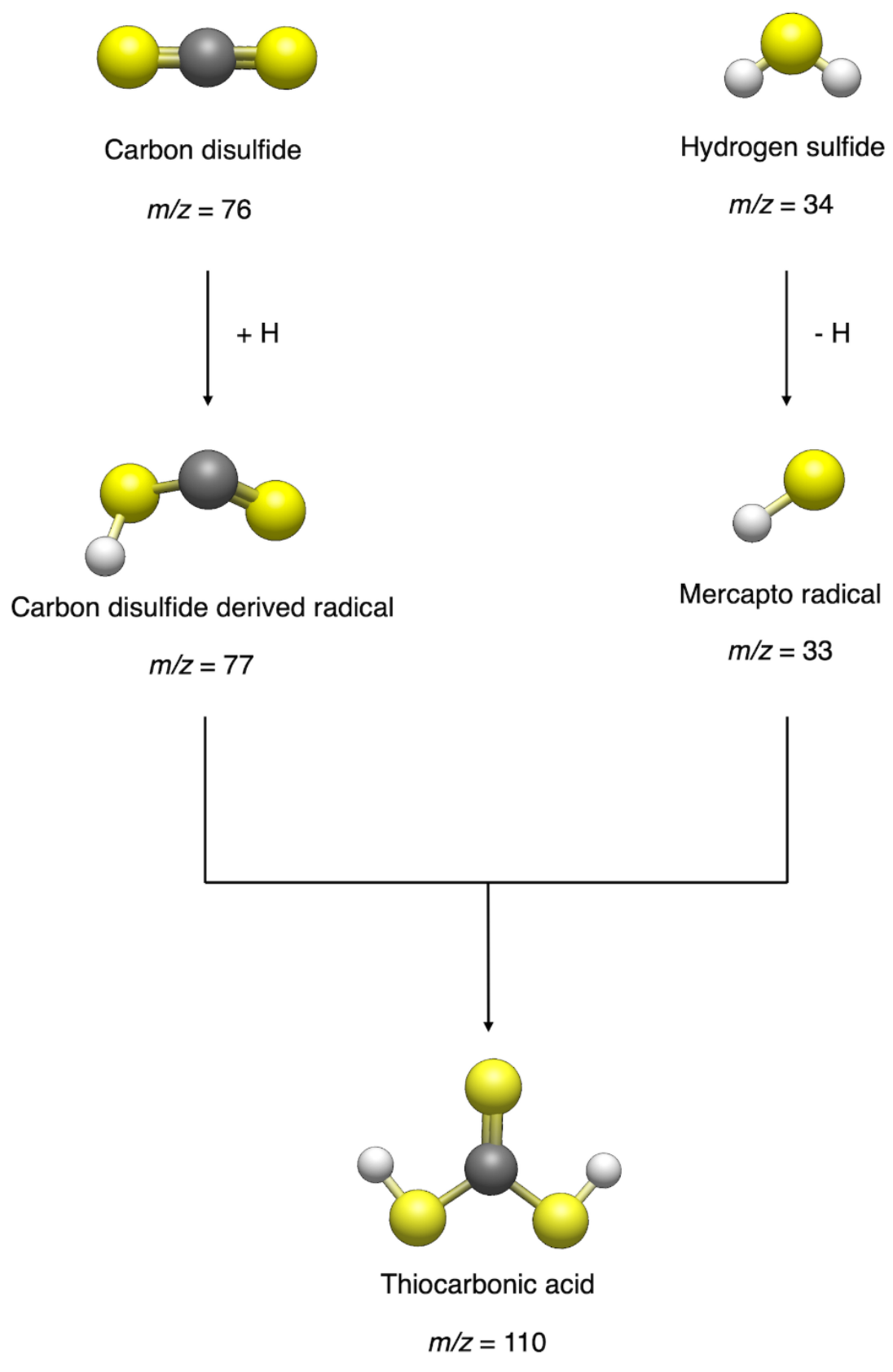

**Scheme 1** Proposed formation pathway of thiocarbonic acid (**1**) from hydrogen sulfide and carbon disulfide.

The barrierless radical-radical recombination of the mercapto radical and radical 1 leads to the formation of thiocarbonic acid (**1**) via an exoergic reaction by 280 kJ mol$^{-1}$ (addition [2], Fig. 10A). Since the overall mechanism pathway is endoergic by 33 kJ mol$^{-1}$, irradiation of $H_2S$−$CS_2$ ice by mimicking GCRs is critical for the formation of thiocarbonic acid (**1**). This pathway leading to thiocarbonic acid (**1**) can be compared to a potential pathway to undetected isomer **2** as proposed in Figure 10B. The formation of isomer **2** also starts with the formation of a mercapto radical followed by the addition of Ḣ onto carbon disulfide, leading to the HC(S)Ṡ radical (radical 2). The reaction to form HC(S)Ṡ radical has an activation energy of 36 kJ mol$^{-1}$ and is exoergic by 94 kJ mol$^{-1}$ (addition [1], Fig. 10B). The barrierless radical-radical recombination of the mercapto radical and radical 2 leads to isomer **2** through an exoergic reaction by 239 kJ mol$^{-1}$ (addition [2], Fig. 10A). Overall, the formation of isomer **2** is endoergic by 45 kJ mol$^{-1}$. Radical 2 is thermodynamically more stable than radical 1 and its formation is more exoergic than that of radical 1. However, the smaller entrance barrier for addition [1] indicates the preferential formation of radical 1, which forms isomer **1**. The activation energy from radical 1 to form radical 2 is calculated to be as high as 117 kJ mol$^{-1}$ (Fig. S5), which indicates the interconversion is unlikely to proceed. Additionally, the formation of isomer **2** is endoergic by 12 kJ mol$^{-1}$ more than that of isomer **1**, suggesting the preferential formation of thiocarbonic acid (**1**) in irradiated hydrogen sulfide–carbon disulfide ices.

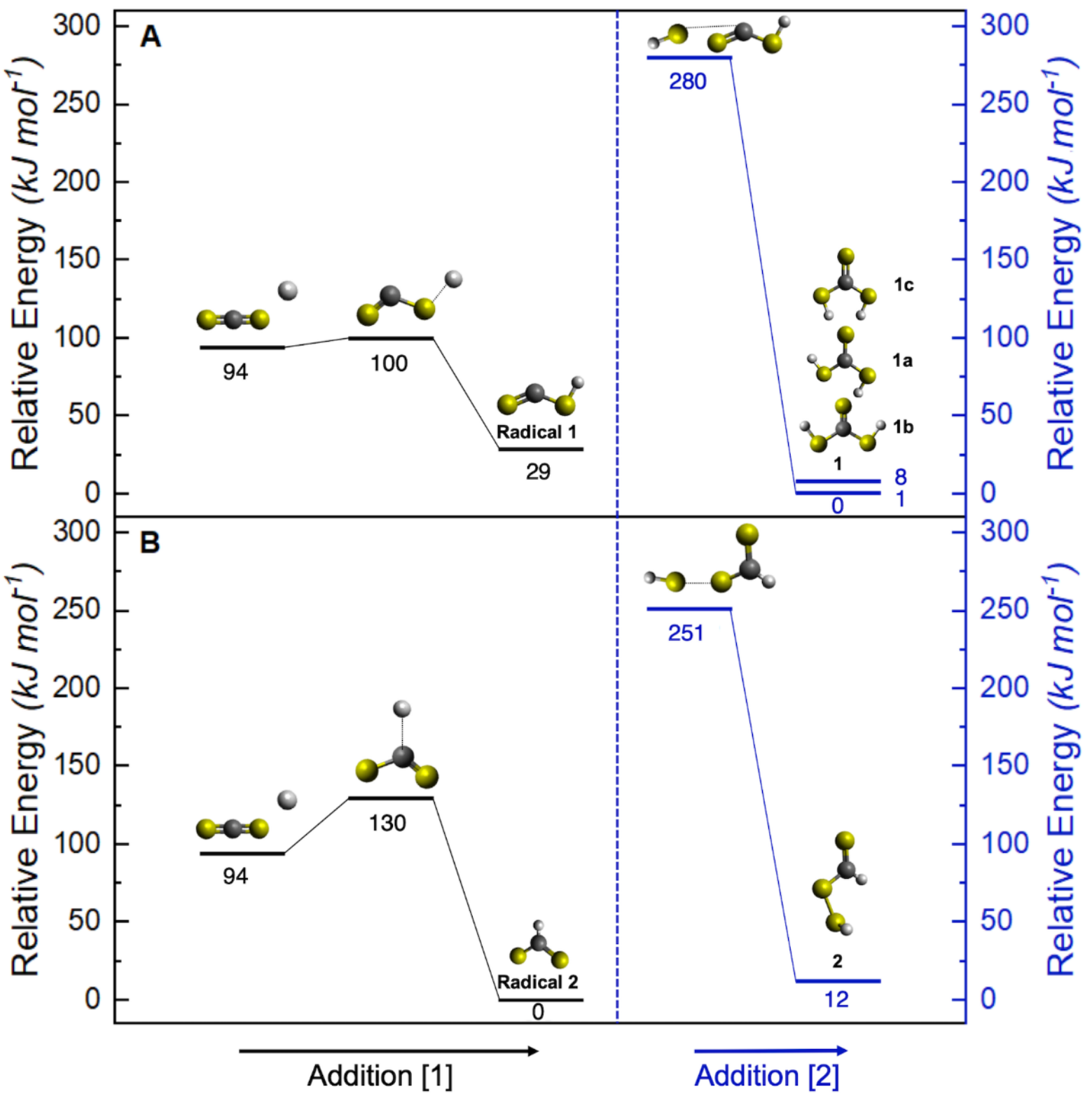

**Fig. 10** Potential energy surface for the proposed reaction mechanism leading to isomer **1** (A) and isomer **2** (B), calculated at the CCSD(T)-F12/cc-pVTZ-F12 level of theory. Addition [1] corresponds to the reaction of a hydrogen atom with $CS_2$ forming radical 1 (mercaptothioxo, HSĊS). Addition [2] corresponds to the barrierless radical-radical recombination between radical 1 (A) or radical 2 (HC(S)Ṡ) (B) and the mercapto radical (ṠH). The formation of the mercapto radical is not represented on the potential energy surface for clarity. Hydrogen, carbon, and sulfur atoms are represented by white, black and yellow spheres respectively.

## Conclusion

Our results present the first experimental study of the formation of thiocarbonic acid (**1**) in interstellar ice analogs composed of hydrogen sulfide ($H_2S$) and carbon disulfide ($CS_2$). The ice mixtures were exposed to electron irradiation simulating the impact of GCRs on interstellar ices. Laboratory analysis using PI-ReToF-MS, combined with computational studies, provide insights into the formation pathway of thiocarbonic acid (**1**) through the barrierless radical-radical recombination of mercaptothioxo (HSĊS) and mercapto (ṠH) radicals. Thiocarbonic acid (**1**) was identified in the gas phase isomer selectively during the TPD process in form of monomers and dimers. These results highlight the unique ability of PI-ReToF-MS to identify complex sulfur-containing molecules in laboratory simulation experiments.

Although thiocarbonic acid (**1**) has not been detected via astronomical observation, both reactants $H_2S$ and $CS_2$ have been observed in the gas phase of various interstellar environments.[11–18] Before being released into the gas phase in star forming regions, these species are likely present in the interstellar ices and subjected to ionizing sources such as GCRs, leading to the formation of thiocarbonic acid (**1**). Once formed, thiocarbonic acid (**1**) may react with other species adsorbed on the ice mantle, contributing to the formation of more complex sulfur-bearing molecules and playing a crucial role in the sulfur depletion observed in the ISM.[3] During protostar formation, the warming of these ices can lead to the desorption of thiocarbonic acid (**1**) into the gas phase, making its detection possible via radio telescopes such as the Atacama Large Millimeter/submillimeter Array (ALMA). To the best of our knowledge, no experimental measurement of the rotational spectrum of thiocarbonic acid (**1**) has been reported. Therefore, a crucial next step toward its potential identification in the ISM would be producing such spectroscopic data.

Future experiments could explore the formation of other carbonic acid related molecules containing both oxygen and sulfur atoms such as carbonothioic acid (HSC(O)OH) and carbonodithioic acid (HSC(O)SH) using ice mixtures containing $H_2O$ and OCS. Since $H_2O$ is a major component of icy dust grains and OCS is one of the two S-bearing molecules detected in the solid phase,[8] such studies could provide valuable insights into the sulfur depletion problem.[3]

## Author contributions

L.C. : Conceptualization, Formal analysis, Investigation, Methodology, Project administration, Visualization, Writing – original draft, Writing – review & editing

J.W. : Investigation, Writing – review & editing

A.H. : Investigation

A.M.T. : Conceptualization, Project administration, Supervision

M.M. : Isotopically labeled ($H_2S$–$^{13}CS_2$) investigation

R.C.F. : Data curation, Funding acquisition, Investigation, Writing – review & editing

R.I.K. : Conceptualization, Funding acquisition, Project administration, Supervision, Writing – review & editing

## Conflicts of Interest

There are no conflicts to declare.

## Acknowledgements

We acknowledge support from the U.S. National Science Foundation (NSF), Division of Astronomical Sciences (AST2403867), awarded to the University of Hawaii at Manoa (R.I.K). We also thank the W. M. Keck Foundation and the University of Hawaii at Manoa for their support in constructing the experimental setup. R.C.F. wishes to thank NASA grant 80NSSC24M0132 and NSF grant AST-2407815 as well as computing resources provided by the Mississippi Center for Supercomputing Research.

# Supporting Information for

## Formation of Thiocarbonic Acid ($H_2CS_3$) – the Sulfur Counterpart of Carbonic Acid ($H_2CO_3$) – in Interstellar Analog Ices

Lina Coulaud,[1,2†] Jia Wang,[1,2] Ashanie Herath,[1,2] Andrew M. Turner,[1,2] Mason Mcanally,[1,2] Ryan C. Fortenberry,[3*] Ralf I. Kaiser[1,2*]

[1] W. M. Keck Research Laboratory in Astrochemistry, University of Hawaii at Manoa, Honolulu, HI 96822, USA

[2] Department of Chemistry, University of Hawaii at Manoa, Honolulu, HI 96822, USA

[3] Department of Chemistry & Biochemistry, University of Mississippi, MS 38677, USA

[†] Present address: Département de Chimie, Université Paris-Saclay, Gif-sur-Yvette, 91190, France

[*]Corresponding Author: r410@olemiss.edu, ralfk@hawaii.edu

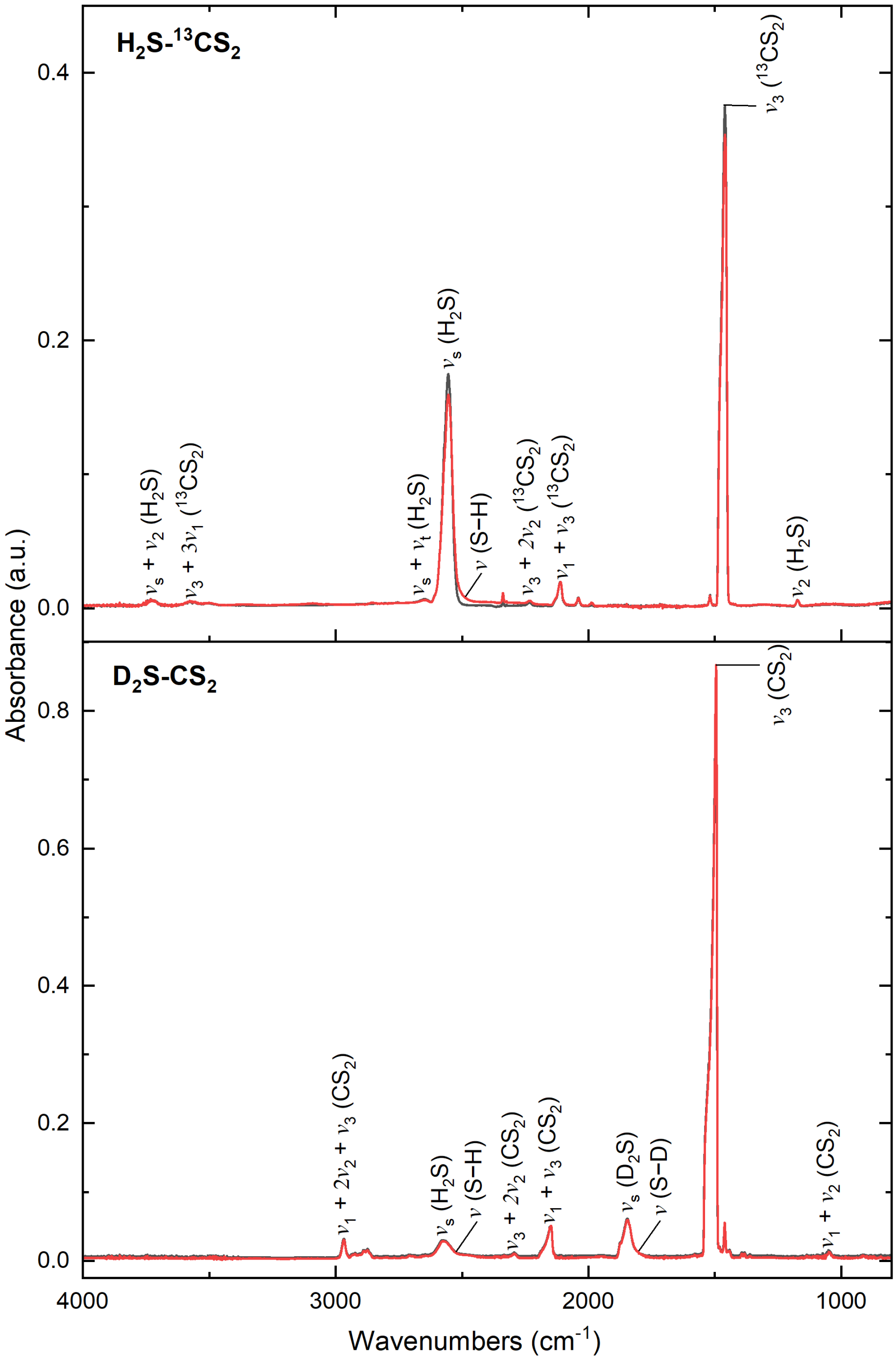

**Fig. S1** IR spectrum of pristine (black) and irradiated (red) 6.1:1 $H_2S-^{13}CS_2$ ice (top) and 0.5:1 $D_2S-CS_2$ ice (bottom) at 5 K. Assignments were given based on detailed wavenumbers listed in Table S3.

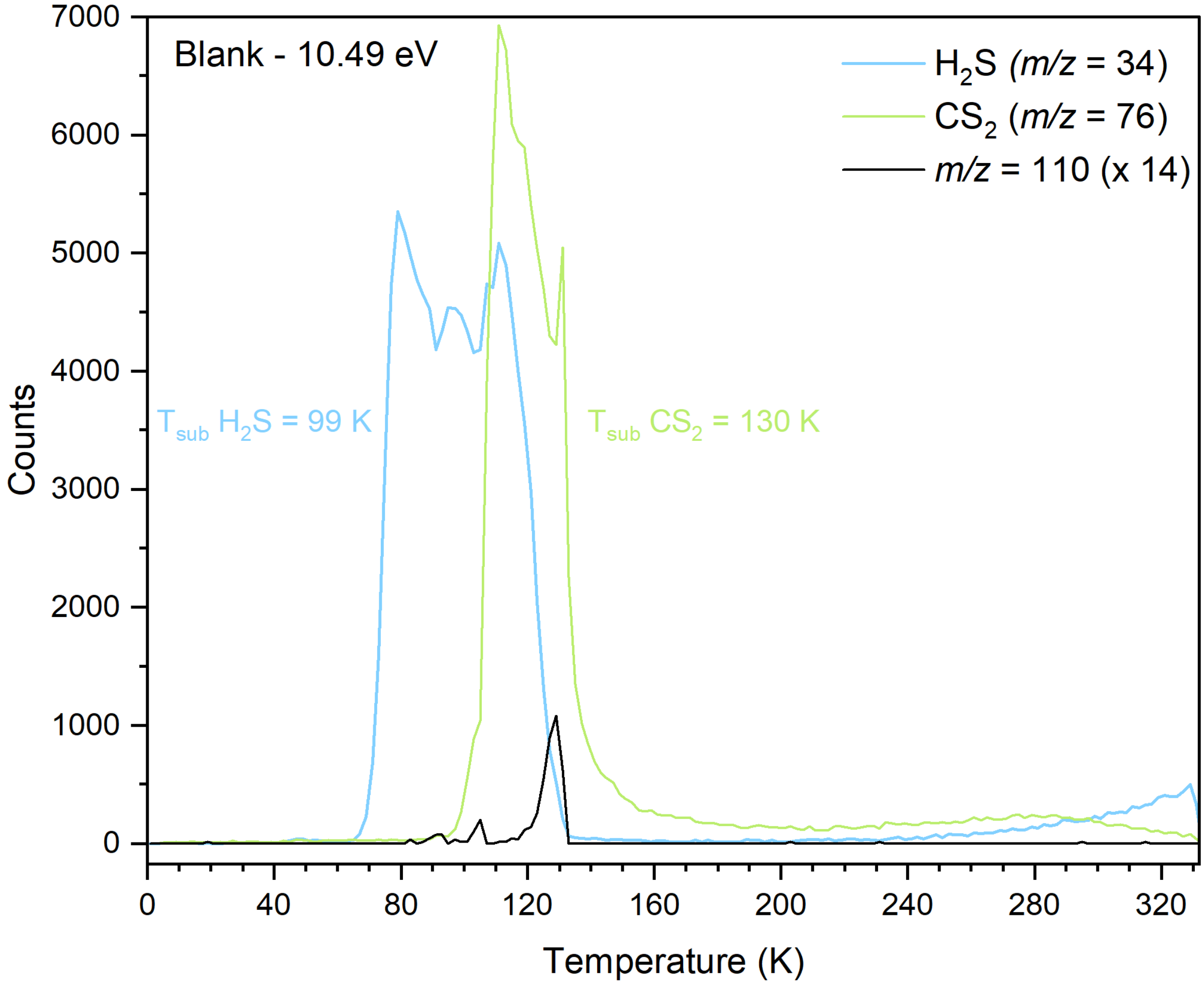

**Fig. S2** TPD profiles of hydrogen sulfide (blue) at *m/z* = 34, carbon disulfide (green) at *m/z* = 76 and the co-desorption features at *m/z* = 110 (black) in unirradiated (blank) $H_2S$−$CS_2$ ice at 10.49 eV. For the TPD profiles of hydrogen sulfide and carbon disulfide, the unusual shape is caused by the saturation of the detector upon sublimation.

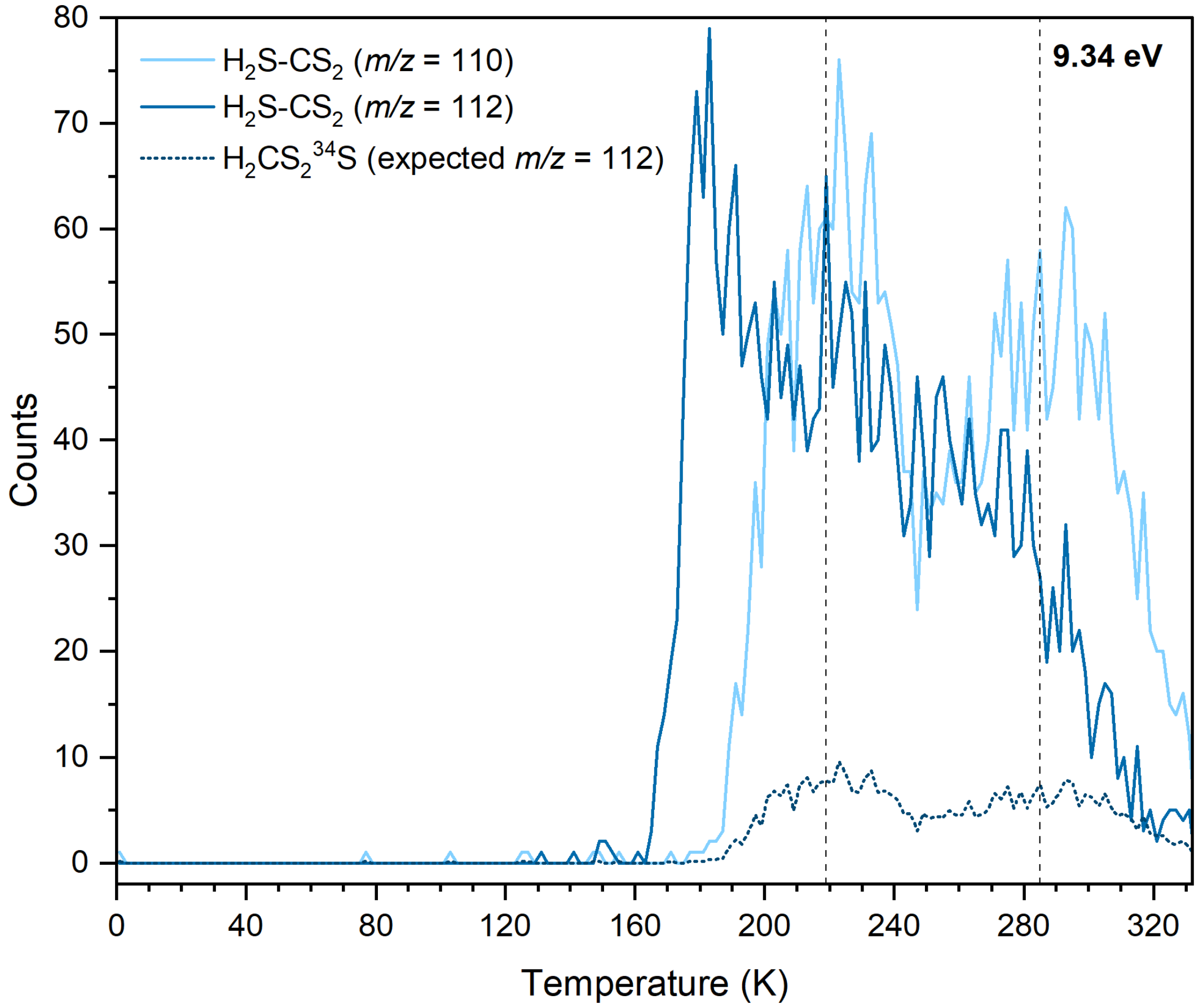

**Fig. S3** TPD profiles at 9.34 eV for : $H_2S{-}CS_2$ (*m/z* = 110), $H_2S{-}CS_2$ (*m/z* = 112), and expected signal for $H_2CS_2{}^{34}S$ (*m/z* = 112). The dashed lines indicate sublimation peaks II (219 K) and III (285 K).

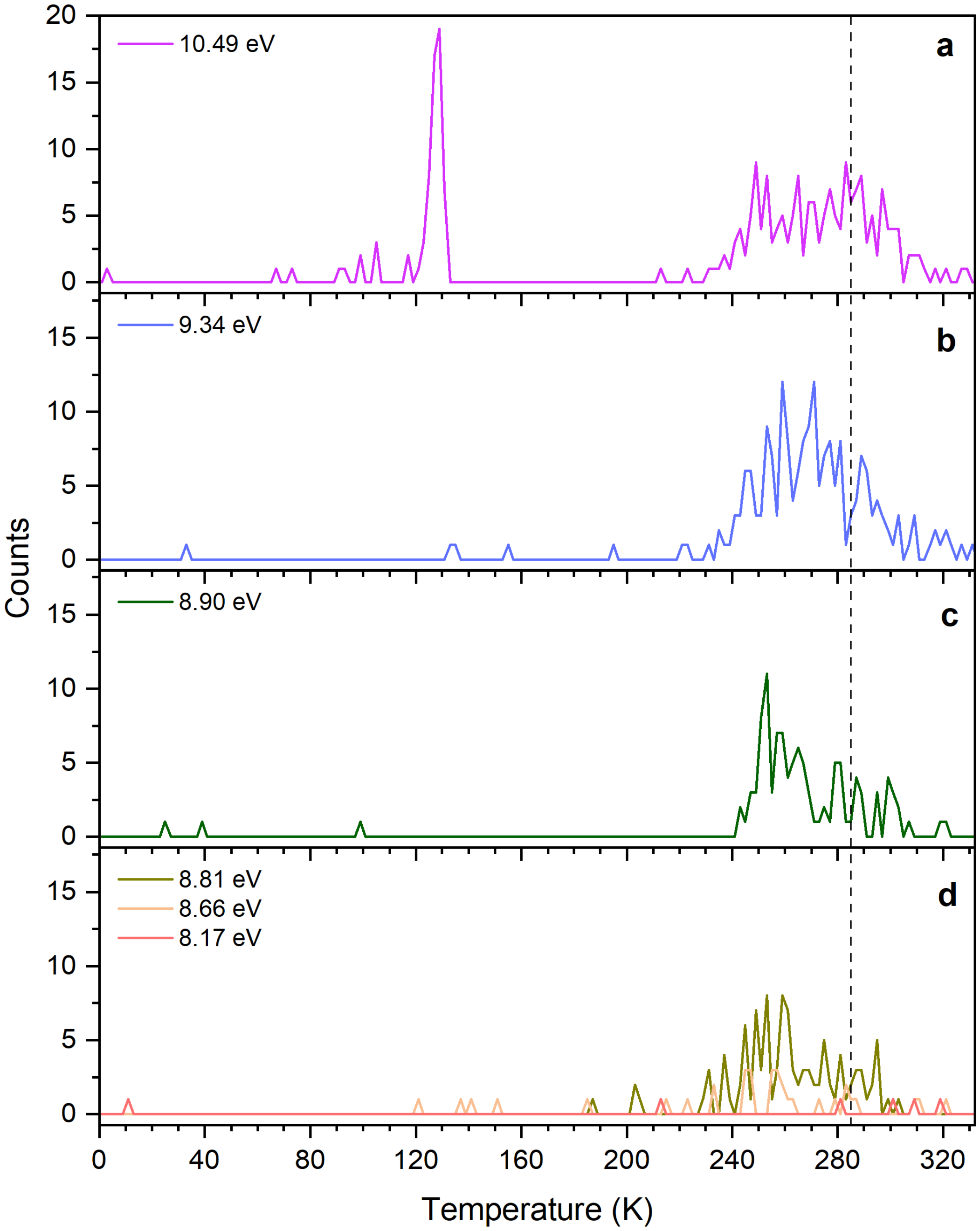

**Fig. S4** TPD profiles of *m/z* = 220 from the irradiated $H_2S-CS_2$ ice at (a) 10.49 eV, (b) 9.34 eV, (c) 8.90 eV, (d) 8.81, 8.66, and 8.17 eV. The dashed black line indicates the sublimation peak at 285 K.

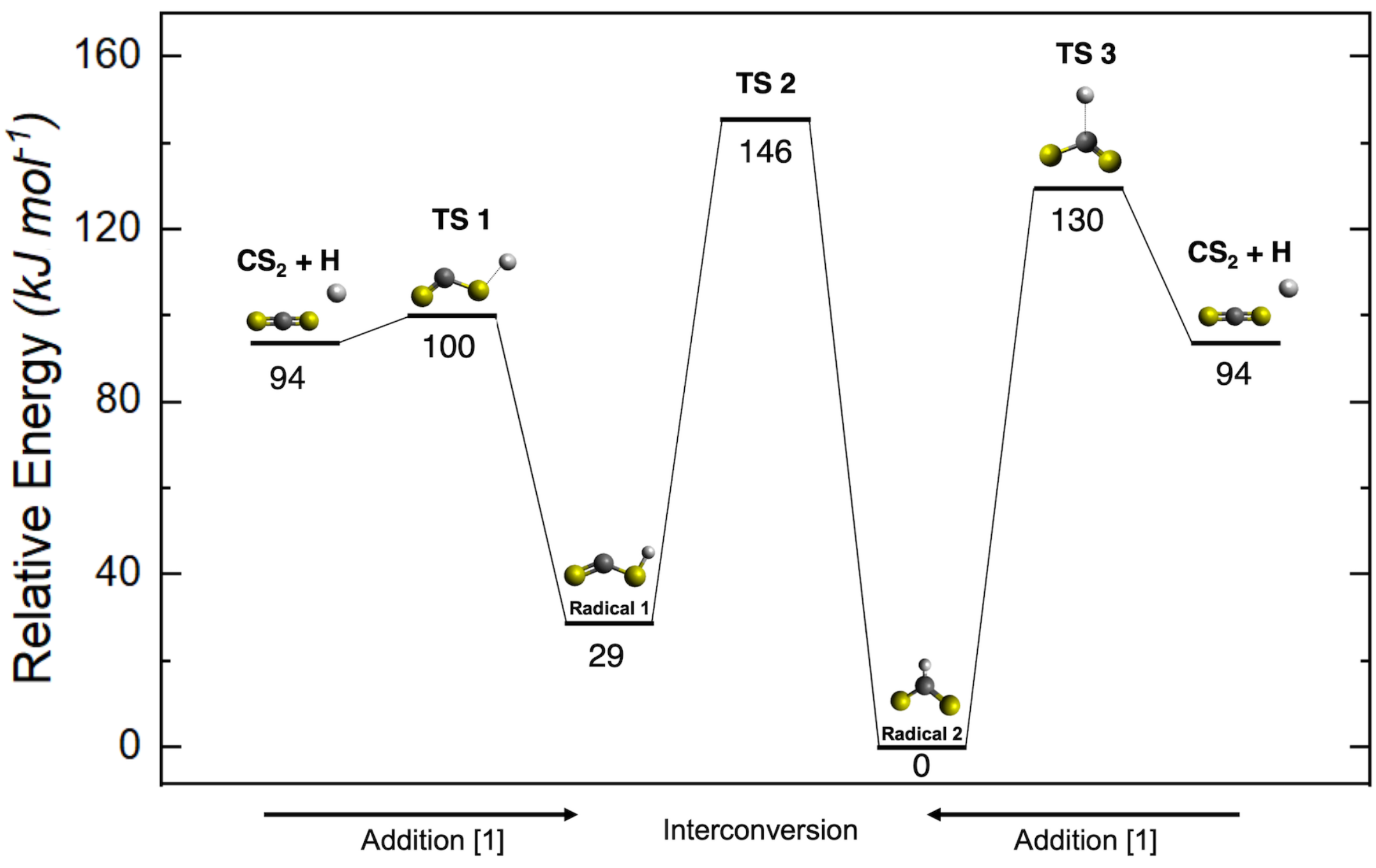

**Fig. S5** CCSD(T)-F12/cc-pVTZ-F12 potential energy surface for the formation and interconversion of radical 1 (HSĊS) and radical 2 (HC(S)Ṡ).

**Table S1** Experimental parameters of the ices : composition, ratio, thickness, irradiation dose (current and irradiation time), and photon energy used during TPD.

| Experiment | Composition | Ratio | Thickness (nm) | Current (nA) | Time (min) | Dose, $H_2S$ (eV molecule$^{-1}$) | Dose, $CS_2$ (eV molecule$^{-1}$) | Photon energy (eV) |
|---|---|---|---|---|---|---|---|---|
| 1 | $H_2S:CS_2$ | 2.7 : 1 | 850 ± 50 | 0 | 0 | 0 | 0 | 10.49 |
| 2 | $H_2S:CS_2$ | 3.7 : 1 | 850 ± 50 | 20 ± 1 | 10 | 0.24 | 0.83 | 10.49 |
| 3 | $H_2S:CS_2$ | 3.5 : 1 | 850 ± 50 | 20 ± 1 | 10 | 0.24 | 0.78 | 9.34 |
| 4 | $H_2S:CS_2$ | 3.5 : 1 | 850 ± 50 | 20 ± 1 | 10 | 0.25 | 0.83 | 8.17 |
| 5 | $H_2S:CS_2$ | 4.1 : 1 | 850 ± 50 | 20 ± 1 | 10 | 0.25 | 0.96 | 8.90 |
| 6 | $H_2S:CS_2$ | 3.7 : 1 | 850 ± 50 | 20 ± 1 | 10 | 0.23 | 0.79 | 8.66 |
| 7 | $H_2S:CS_2$ | 4.3 : 1 | 850 ± 50 | 20 ± 1 | 10 | 0.25 | 1.0 | 8.81 |
| 8 | $D_2S:CS_2$ | 0.5 : 1 | 850 ± 50 | 20 ± 1 | 10 | 1.9 | 0.94 | 9.34 |
| 9 | $H_2S:{}^{13}CS_2$ | 6.1 : 1 | 850 ± 50 | 20 ± 1 | 10 | 0.38 | 2.2 | 9.34 |

**Table S2** Parameters for the generation of vacuum ultraviolet (VUV) photons via four-wave mixing schemes.

| Experiment | Medium | $\omega_{VUV}$ | YAG 1 Wavelength (nm) | $\omega_1$ Dye | $\omega_1$ (nm) | YAG 2 Wavelength (nm) | $\omega_2$ Dye | $\omega_2$ (nm) | Energy (eV) |
|---|---|---|---|---|---|---|---|---|---|
| 1, 2 | Xenon | $3\omega_1$ | 355 | - | 355[a] | - | - | - | 10.49 |
| 3, 8, 9 | Krypton | $2\omega_1 - \omega_2$ | 355 | Stilbene 420 | 212.556 | 532 | - | 532[b] | 9.34 |
| 4 | Krypton | $2\omega_1 - \omega_2$ | 355 | Stilbene 420 | 212.556 | 355 | - | 355[a] | 8.17 |
| 5 | Krypton | $2\omega_1 - \omega_2$ | 355 | Stilbene 420 | 212.556 | 355 | Coumarin 450 | 448.239 | 8.90 |
| 6 | Xenon | $2\omega_1 - \omega_2$ | 355 | Coumarin 450 | 222.566 | 355 | Coumarin 503 | 500 | 8.66 |
| 7 | Xenon | $2\omega_1 - \omega_2$ | 355 | Coumarin 450 | 222.566 | 532 | - | 532[b] | 8.81 |

[a]Nd :YAG third harmonic
[b]Nd :YAG second harmonic

**Table S3** Absorption peaks observed in $H_2S$−$CS_2$ ice and isotopic mixtures before and after electron irradiation at 5 K.

| $H_2S$ ($cm^{-1}$) | $D_2S$ ($cm^{-1}$) | $CS_2$ ($cm^{-1}$) | $^{13}CS_2$ ($cm^{-1}$) | Assignment | Reference |
|---|---|---|---|---|---|
| 3721 | | | | $\nu_s + \nu_2$ ($H_2S$) | [50], [51] |
| | | | 3570 | $\nu_3 + 3\nu_1$ ($^{13}CS_2$) | - |
| | | 3485 | | $\nu_3 + 3\nu_1$ ($CS_2$) | [52] |
| | | 2967 | | $\nu_1 + 2\nu_2 + \nu_3$ ($CS_2$) | [52], [53] |
| 2644 | | | | $\nu_s + \nu_l$ ($H_2S$) | [50], [51] |
| 2546 | | | | $\nu_s$ ($\nu_1$ and $\nu_3$) ($H_2S$) | [50], [51] |
| | | 2295 | | $\nu_3 + 2\nu_2$ ($CS_2$) | [52] |
| | | | 2232 | $\nu_3 + 2\nu_2$ ($^{13}CS_2$) | - |
| | | 2153 | | $\nu_1 + \nu_3$ ($CS_2$) | [52], [53] |
| | | | 2112 | $\nu_1 + \nu_3$ ($^{13}CS_2$) | [54] |
| | 1847 | | | $\nu_s$ ($\nu_1$ and $\nu_3$) ($D_2S$) | [51] |
| | | 1499 | | $\nu_3$ ($CS_2$) | [52],[53],[54] |
| | | | 1460 | $\nu_3$ ($^{13}CS_2$) | [54] |
| 1166 | | | | $\nu_2$ ($H_2S$) | [50], [51] |
| | | 1043 | | $\nu_1 + \nu_2$ ($CS_2$) | [52] |

**Table S4** Adiabatic ionization energies (IEs) and relative energies (Rel. E) of thiocarbonic acid dimers ($H_4C_2S_6$) ($m/z$ = 220) were computed at the CCSD(T)-F12/cc-pVTZ-F12 + ZPVE(ωB97XD/aug-cc-pVTZ) level of theory. The calculated energies were corrected for the combined error limits of ± 0.04 eV and the thermal and Stark effect by −0.03 eV.

| | Conformer | Structure | Rel. E (kJ/mol) | IE (eV) | IE range with error (eV) | Corrected IE with Stark effect (eV) |
|---|---|---|---|---|---|---|
| **1a-1a** | ct-ct | 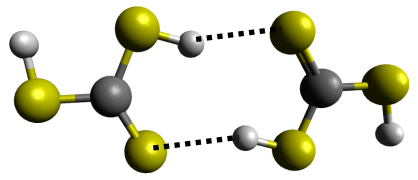 | 1 | 7.93 | 7.89–7.97 | 7.86–7.94 |
| **1b-1a** | cc-ct | 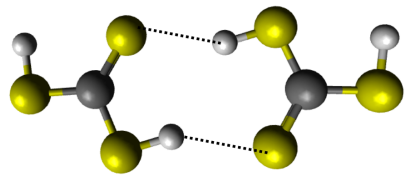 | 0 | 7.93 | 7.89–7.97 | 7.86–7.94 |
| **1b-1b** | cc-cc | 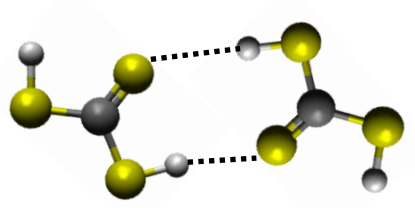 | 0 | 8.93 | 8.89–8.97 | 8.86–8.94 |
| **1c-1c** | tt-tt | 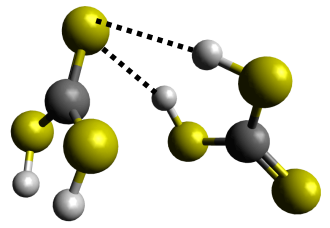 | 14 | 8.32 | 8.28–8.36 | 8.25–8.33 |

**Table S5 CCSD(T)-F12/cc-pVTZ-F12 optimized geometries (in Å).**

Isomer 1b+
CCSD(T)-F12/CC-PVTZ-F12 ENERGY=-1232.19956925

| | | | |
|---|---|---|---|
| S | 0.0000000000 | 0.0000000000 | -1.7145548140 |
| C | 0.0000000000 | 0.0000000000 | -0.0439406003 |
| S | 0.0000000000 | 1.4492020350 | 0.8711853682 |
| S | 0.0000000000 | -1.4492020350 | 0.8711853682 |
| H | 0.0000000000 | 2.2902587959 | -0.1805702321 |
| H | 0.0000000000 | -2.2902587959 | -0.1805702321 |

--
Isomer 1b
CCSD(T)-F12/CC-PVTZ-F12 ENERGY=-1232.52616031

| | | | |
|---|---|---|---|
| S | 0.0000000000 | 0.0000000000 | -1.7432588191 |
| C | 0.0000000000 | 0.0000000000 | -0.1138924571 |
| S | 0.0000000000 | 1.4353460125 | 0.8972765856 |
| S | 0.0000000000 | -1.4353460125 | 0.8972765856 |
| H | 0.0000000000 | 2.2870389749 | -0.1371781177 |
| H | 0.0000000000 | -2.2870389749 | -0.1371781177 |

--
Isomer 1a+
CCSD(T)-F12/CC-PVTZ-F12 ENERGY=-1232.19889542

| | | | |
|---|---|---|---|
| S | 0.0000000000 | 1.6948034708 | -0.1714993834 |
| C | 0.0000000000 | 0.0273015591 | 0.0021159258 |
| S | 0.0000000000 | -0.6420152461 | 1.5749912679 |
| S | 0.0000000000 | -1.0014060275 | -1.3687275219 |
| H | 0.0000000000 | -1.9281281904 | 1.1734458466 |
| H | 0.0000000000 | -0.0315442791 | -2.3044257168 |

--
Isomer 1a
CCSD(T)-F12/CC-PVTZ-F12 ENERGY=-1232.52586362

| | | | |
|---|---|---|---|
| S | 0.0000000000 | -0.8505122958 | -1.5045643799 |

C 0.0000000000 -0.0599788464 -0.0805845634
S 0.0000000000 1.6894984492 -0.0424168882
S 0.0000000000 -0.8095794868 1.5040147091
H 0.0000000000 1.8447624359 1.2927365261
H 0.0000000000 -2.0653825182 1.0341966965
--
Isomer 1c+
CCSD(T)-F12/CC-PVTZ-F12 ENERGY=-1232.19509331
S 0.0000000000 0.0000000000 -1.6763592443
C 0.0000000000 0.0000000000 0.0122769986
S 0.0000000000 1.5330832235 0.7718421125
S 0.0000000000 -1.5330832235 0.7718421125
H 0.0000000000 1.0787894060 2.0368782293
H 0.0000000000 -1.0787894060 2.0368782293
--
Isomer 1c
CCSD(T)-F12/CC-PVTZ-F12 ENERGY=-1232.52260092
S 0.0000000000 0.0000000000 -1.6952802937
C 0.0000000000 0.0000000000 -0.0644317538
S 0.0000000000 1.5233731627 0.7951612429
S 0.0000000000 -1.5233731627 0.7951612429
H 0.0000000000 1.0573915280 2.0531168102
H 0.0000000000 -1.0573915280 2.0531168102
--
Isomer 2
CCSD(T)-F12/CC-PVTZ-F12 ENERGY=-1232.52287668
C -0.4994182041 0.0105456054 -0.7998519934
S 0.7314460846 -0.0006957421 0.4333527124
S -0.3512915563 0.0406688955 2.1801167605
H -0.4400204307 -1.2868863505 2.3616495903
S -0.1315482009 -0.0045412512 -2.3752388730
H -1.5162535602 0.0342224531 -0.4078137881

--
Isomer 2+
CCSD(T)-F12/CC-PVTZ-F12 ENERGY=-1232.19269359

| | | | |
|---|---|---|---|
| C | -0.0312986086 | 0.4248265444 | -0.7692794964 |
| S | 0.0017504368 | -0.7434498963 | 0.4496916159 |
| S | 0.0460232672 | 0.3795060987 | 2.1673957052 |
| H | -1.2646756719 | 0.2260241880 | 2.4306608034 |
| S | 0.0057996201 | 0.1513708207 | -2.3909553884 |
| H | -0.0663888595 | 1.4729837058 | -0.4563108767 |

--
Isomer 3
CCSD(T)-F12/CC-PVTZ-F12 ENERGY=-1232.51419855

| | | | |
|---|---|---|---|
| C | -0.6341251965 | 0.0552773704 | 0.2548849922 |
| S | 0.1473157026 | -0.0018935805 | 1.8589566590 |
| H | -0.4675843834 | 1.0875855125 | 2.3441496827 |
| S | 0.0610465097 | -1.0675413969 | -0.9513349380 |
| S | 0.0978918383 | 1.0119183054 | -1.0852698020 |
| H | -1.7170963790 | 0.0831667209 | 0.2690751480 |

--
Isomer 3+
UCCSD(T)-F12/CC-PVTZ-F12 ENERGY=-1232.20148490

| | | | |
|---|---|---|---|
| H | -1.7253041733 | 0.0000000000 | 0.3272367236 |
| C | -0.6399171524 | 0.0000000000 | 0.3089378699 |
| S | 0.0984193555 | 0.0000000000 | 1.9014232601 |
| H | 1.3681913560 | 0.0000000000 | 1.4604537018 |
| S | 0.0762736849 | -0.9952431134 | -1.0366838297 |
| S | 0.0762736849 | 0.9952431134 | -1.0366838297 |

--
Isomer 4
CCSD(T)-F12/CC-PVTZ-F12 ENERGY=-1232.51226316

| | | | |
|---|---|---|---|
| S | 0.1337369104 | -1.2516685987 | 0.0000000000 |
| C | 0.4100442389 | 1.3569833922 | 0.0000000000 |
| H | 1.4965535728 | 1.4385392904 | 0.0000000000 |
| H | -0.0676580399 | 2.3341323277 | 0.0000000000 |
| S | -0.1661398575 | 0.3123384454 | 1.3823403838 |
| S | -0.1661398575 | 0.3123384454 | -1.3823403838 |

--

Isomer 4+

CCSD(T)-F12/CC-PVTZ-F12 ENERGY=-1232.21813879

| | | | |
|---|---|---|---|
| S | 0.0000000000 | 0.0000000000 | -1.2071086634 |
| C | 0.0000000000 | 0.0000000000 | 1.4314023631 |
| H | -0.9005760448 | 0.0000000000 | 2.0465772197 |
| H | 0.9005760448 | 0.0000000000 | 2.0465772197 |
| S | 0.0000000000 | 1.4028232890 | 0.2710804099 |
| S | 0.0000000000 | -1.4028232890 | 0.2710804099 |

**Table S6 CCSD(T)-F12/cc-pVTZ-F12 harmonic frequencies (in cm$^{-1}$).**

Isomer 1b+

| | |
|---|---|
| 1 | 77.08 |
| 2 | 227.86 |
| 3 | 247.87 |
| 4 | 385.58 |
| 5 | 470.20 |
| 6 | 519.06 |
| 7 | 842.63 |
| 8 | 876.13 |
| 9 | 1037.24 |
| 10 | 1140.98 |
| 11 | 2658.45 |
| 12 | 2658.60 |

--

Isomer 1b

| | |
|---|---|
| 1 | 212.54 |
| 2 | 253.94 |
| 3 | 295.35 |
| 4 | 406.83 |
| 5 | 464.33 |
| 6 | 504.56 |
| 7 | 765.63 |
| 8 | 915.60 |
| 9 | 997.34 |
| 10 | 1170.62 |
| 11 | 2705.82 |
| 12 | 2707.50 |

--

Isomer 1a+

| | |
|---|---|
| 1 | 146.35 |
| 2 | 233.82 |
| 3 | 255.23 |
| 4 | 376.16 |
| 5 | 465.68 |
| 6 | 517.27 |
| 7 | 837.15 |
| 8 | 896.73 |
| 9 | 1029.51 |
| 10 | 1137.00 |
| 11 | 2650.29 |
| 12 | 2654.56 |

--

Isomer 1a

| | |
|---|---|
| 1 | 245.88 |
| 2 | 256.38 |
| 3 | 280.47 |
| 4 | 370.86 |
| 5 | 468.54 |
| 6 | 499.46 |
| 7 | 815.22 |
| 8 | 901.21 |
| 9 | 1021.29 |
| 10 | 1170.04 |
| 11 | 2671.13 |
| 12 | 2699.75 |

--

Isomer 1c

| | |
|---|---|
| 1 | 143.78 |
| 2 | 259.18 |
| 3 | 284.05 |
| 4 | 286.24 |
| 5 | 475.64 |
| 6 | 489.53 |
| 7 | 833.34 |
| 8 | 927.54 |
| 9 | 1013.56 |
| 10 | 1157.49 |
| 11 | 2682.47 |
| 12 | 2698.92 |

--

Isomer 2

| | |
|---|---|
| 1 | 116.21 |
| 2 | 171.45 |
| 3 | 303.60 |
| 4 | 325.14 |
| 5 | 528.67 |
| 6 | 770.41 |
| 7 | 807.62 |
| 8 | 894.45 |
| 9 | 1068.81 |
| 10 | 1253.18 |
| 11 | 2674.23 |
| 12 | 3100.85 |

--

Isomer 2+

| | |
|---|---|
| 1 | 129.75 |
| 2 | 173.02 |
| 3 | 277.96 |
| 4 | 296.50 |
| 5 | 528.11 |
| 6 | 763.67 |
| 7 | 796.76 |
| 8 | 903.48 |
| 9 | 964.13 |
| 10 | 1170.54 |
| 11 | 2667.93 |
| 12 | 3072.66 |

--

Isomer 3

| | |
|---|---|
| 1 | 210.67 |
| 2 | 235.45 |
| 3 | 285.75 |
| 4 | 510.94 |
| 5 | 588.66 |
| 6 | 688.70 |
| 7 | 901.49 |
| 8 | 960.65 |
| 9 | 1025.98 |
| 10 | 1229.22 |
| 11 | 2689.14 |
| 12 | 3163.27 |

--

Isomer 3+

1 229.54
2 243.113
3 316.61
4 511.81
5 569.44
6 676.51
7 885.28
8 997.27
9 1028.37
10 1269.94
11 2689.00
12 3176.66
--
Isomer 4

| | |
|---|---|
| 1 | 184.17 |
| 2 | 366.56 |
| 3 | 477.81 |
| 4 | 507.78 |
| 5 | 717.24 |
| 6 | 722.33 |
| 7 | 880.62 |
| 8 | 1080.15 |
| 9 | 1205.70 |
| 10 | 1457.18 |
| 11 | 3064.43 |
| 12 | 3148.31 |

--
Isomer 4+

| | |
|---|---|
| 1 | 125.61 |
| 2 | 385.27 |
| 3 | 524.62 |
| 4 | 568.06 |
| 5 | 713.51 |
| 6 | 762.24 |
| 7 | 795.20 |
| 8 | 1033.63 |
| 9 | 1193.65 |
| 10 | 1426.33 |
| 11 | 3063.89 |
| 12 | 3139.85 |

**Table S7 ωB97XD/aug-cc-pVTZ optimized geometries (in Å).**

1b-1b+
C,0.0279409892,0.0504624027,-0.1365042332
S,0.0667939071,0.2556774992,1.5691441155
H,1.2484636109,-0.3591342705,1.7631233554
S,1.1107735419,-0.8555184921,-0.9997462673
S,-1.2700681065,0.9026898111,-0.8998545459
H,-1.1924817922,0.3026001448,-2.1223943173
H,0.9757172388,-0.1204310858,-3.3044095348
S,1.2626371264,0.3909995791,-4.5361812652
C,-0.2592386589,0.0592062084,-5.2898478109
S,-0.2242192381,0.2345769867,-6.9990205667
H,-1.5495621083,0.0932385437,-7.1872711849
S,-1.5961246484,-0.3725233694,-4.4156432814
--
1b-1b
C,-0.0883001682,0.,0.2690574527
S,-0.0944549542,0.,2.0225195742
H,1.240901691,0.,2.1499132481
S,1.2732659201,0.,-0.6384946517
H,0.8953502126,0.,-3.2008014158
S,-1.7251670985,0.,-0.329533054
H,-1.7597637054,0.,-6.9956526686

S,-0.4244070601,0.,-6.8682589947
C,-0.4305618461,0.,-5.1147968732
S,-1.7921279344,0.,-4.2072447688
H,-1.4142122269,0.,-1.6449380048
S,1.2063050842,0.,-4.5162063665
--
1a-1a+
C,0.0649026198,0.3186385038,-0.4508512476
S,0.0324935955,0.3153995576,1.2704448423
H,1.0832559942,-0.5051918757,1.4673489578
S,-1.0917755796,1.2673648136,-1.2039935598
S,1.2705270292,-0.6387123193,-1.2168980515
H,0.9579292452,-0.3024049949,-2.4837501237
H,-2.3565125743,-0.5252944324,-2.552174163
S,-2.6171341308,-0.9804645676,-3.7934142734
C,-1.5458758945,0.0768884305,-4.6247630534
S,-1.5068249094,-0.032052636,-6.3424642166
H,-2.4388970675,-0.9954508819,-6.481592581
S,-0.52757764,1.2163183234,-3.9391676615
--
1a-1a
C,0.100026044,0.,0.2621752163
S,-0.0303130602,0.,2.0059230083
H,1.2823031748,0.,2.3022705475
S,-1.2719848597,0.,-0.6284185974
H,-0.9069956274,0.,-3.1851199114
S,1.7318864798,0.,-0.3429099567
H,-0.7634411565,0.,-7.1480099674
S,0.5491750784,0.,-6.8516624282
C,0.4188359743,0.,-5.1079146362
S,1.790846878,0.,-4.2173208225
H,1.4258576457,0.,-1.6606195085

S,-1.2130244616,0.,-4.5028294632
--
1c-1c+
C,0.0912036514,-0.0856492217,-0.2221036585
S,0.1672618493,-0.5467770946,1.4017888325
S,1.0984266069,1.2104403885,-0.6940766693
H,0.7209392053,1.2283829451,-2.0172585557
S,-0.9862308463,-0.9644515835,-1.2140551789
H,-0.7231128363,-0.2782758595,-2.3774861658
H,1.9842122922,-0.2758973563,-7.0363621871
S,1.0055236892,0.5819391242,-6.6990784232
C,0.9245908433,0.2585065627,-5.0101205555
S,-0.1904356051,1.1167111361,-4.1223595399
S,1.9295803739,-0.8967756703,-4.2018845895
H,2.656223524,-1.3481533698,-5.2380534281
--
1c-1c
C,0.1772794396,0.5890534913,-1.4664295486
S,0.3643423713,1.8774078116,-0.4834042639
S,1.4790306387,-0.5627488238,-1.6386012269
H,0.9788515486,-1.3389595769,-2.6209793249
S,-1.3384381838,0.381048334,-2.3171701942
H,-1.0970749475,-0.7884356214,-2.9374662544
H,-1.2538905431,1.041364041,-6.6660849072
S,-1.516745015,-0.2435284422,-6.3745890071
C,-0.0039305755,-0.6956884969,-5.6375706589
S,0.1765631214,-2.2309823216,-5.0896075894
S,1.3117045965,0.4290815957,-5.4627997266
H,0.7223075525,1.5423880514,-5.9295103469
--
1b-1a+

C,0.3072158574,0.0230598745,-0.2140299169
S,0.1533786923,0.2840896506,1.4843615875
H,1.1251179306,-0.5687316103,1.8553180549
S,1.4269705864,-1.0586118731,-0.8264120584
S,-0.7995237066,0.9803475099,-1.122142575
H,-0.456788468,0.4914662908,-2.3291973763
H,2.7924083406,0.5479969637,-2.3275035975
S,3.10234199,0.8591929606,-3.6011902747
C,2.019955554,-0.2495752021,-4.3476214764
S,2.0315950249,-0.3244592635,-6.0676607431
H,2.9968473245,0.5917708168,-6.2799656785
S,0.9455706889,-1.2754333189,-3.5751190905
--
1b-1a
C,-0.0061105525,0.,0.0120383898
S,-0.0315432545,0.,1.7655348141
H,1.3023664354,0.,1.9074085694
S,1.3658217295,0.,-0.8799742473
S,-1.6365476986,0.,-0.6032226011
H,-1.3143591095,0.,-1.9158466884
H,1.0143927371,0.,-3.4443441173
S,1.3282841193,0.,-4.7599007317
C,-0.3019313908,0.,-5.3711063681
S,-0.4267864578,0.,-7.1150681201
H,0.8867836388,0.,-7.4072776425
S,-1.6757221794,0.,-4.4839600692
--

**Table S8 ωB97XD/aug-cc-pVTZ harmonic frequencies (in cm$^{-1}$).**

1b-1b+

| | | |
|---|---|---|
| 18.5307 | 29.7458 | 30.0679 |
| 80.8118 | 100.5116 | 142.9459 |
| 265.5364 | 272.2777 | 281.9558 |
| 298.7356 | 304.8669 | 323.9614 |
| 423.2035 | 441.3259 | 491.8949 |
| 492.9076 | 520.6841 | 521.3102 |
| 847.1706 | 859.6900 | 891.6135 |
| 893.2739 | 1038.0228 | 1039.7495 |
| 1124.8123 | 1139.6126 | 2640.3277 |
| 2642.6754 | 2672.8778 | 2673.1062 |

--

1b-1b

| | | |
|---|---|---|
| 8.2305 | 22.2712 | 42.2176 |
| 58.9013 | 67.7510 | 67.8920 |
| 266.7585 | 271.6124 | 290.9944 |
| 292.0502 | 294.3931 | 302.3161 |
| 455.0379 | 467.5630 | 489.2924 |
| 491.4814 | 510.4836 | 512.1090 |
| 834.2916 | 842.4711 | 910.9077 |
| 912.8579 | 1034.2885 | 1047.6164 |
| 1149.3785 | 1177.2484 | 2569.4219 |
| 2581.3890 | 2685.2864 | 2685.3372 |

--

1a-1a+

| | | |
|---|---|---|
| 18.3792 | 27.1749 | 30.6116 |
| 64.7460 | 87.0656 | 87.8176 |
| 240.8414 | 255.9695 | 268.9879 |
| 278.7507 | 294.2423 | 295.4832 |
| 411.9800 | 429.5198 | 486.1727 |
| 487.8027 | 526.2787 | 539.0409 |
| 836.9602 | 857.0105 | 890.2226 |
| 913.6431 | 1018.7690 | 1032.2962 |
| 1135.4775 | 1526.7386 | 2441.9734 |
| 2685.5869 | 2690.0285 | 3145.4012 |

--

1a-1a

| | | |
|---|---|---|
| 4.7446 | 20.3643 | 39.6565 |
| 58.8847 | 67.3269 | 68.1730 |
| 265.7640 | 269.4264 | 278.4840 |
| 291.6400 | 302.6775 | 307.5393 |
| 468.5227 | 476.2556 | 485.7182 |
| 491.4223 | 515.3748 | 517.2369 |
| 789.4682 | 798.9745 | 925.3830 |
| 929.4426 | 1015.6072 | 1027.3306 |
| 1154.1340 | 1181.5233 | 2583.0586 |
| 2594.0960 | 2716.3113 | 2716.3233 |

--

1c-1c+

| | | |
|---|---|---|
| 13.9973 | 24.9560 | 33.4417 |
| 54.2421 | 61.8702 | 100.8790 |
| 233.3201 | 241.7456 | 279.0486 |
| 292.9059 | 301.5336 | 350.5808 |
| 406.0829 | 475.5232 | 507.7207 |
| 510.6439 | 523.3673 | 563.6742 |
| 865.2415 | 885.0992 | 901.4396 |
| 955.0570 | 1030.0992 | 1044.6865 |
| 1137.2777 | 1164.8096 | 2211.0755 |
| 2337.6638 | 2695.3091 | 2710.5737 |

--

1c-1c

| | | |
|---|---|---|
| 21.7823 | 28.1278 | 33.3438 |
| 59.9063 | 69.0598 | 76.3137 |
| 206.6954 | 266.4122 | 268.3040 |
| 276.5791 | 293.1274 | 308.6826 |
| 324.0383 | 368.6651 | 494.0415 |
| 498.3774 | 500.5160 | 510.2012 |
| 845.1309 | 850.3192 | 921.7841 |
| 942.7146 | 1025.5307 | 1026.6727 |
| 1149.7002 | 1162.0264 | 2633.4153 |
| 2680.4334 | 2698.8155 | 2714.7402 |
| -- | | |
| 1b-1a+ | | |
| 18.7213 | 29.0706 | 32.2258 |
| 79.3278 | 98.5487 | 145.5564 |
| 261.9690 | 269.5592 | 273.7202 |
| 284.7491 | 302.4987 | 327.0370 |
| 427.8314 | 458.7572 | 491.5881 |
| 496.0215 | 520.8561 | 522.8622 |
| 830.4714 | 851.4062 | 889.5578 |
| 895.2007 | 1034.0504 | 1038.2910 |
| 1127.4932 | 1143.7305 | 2644.9978 |
| 2648.9123 | 2676.6184 | 2692.2880 |
| -- | | |
| 1b-1a | | |
| 7.5003 | 21.1930 | 40.8970 |
| 58.9805 | 67.1149 | 68.3629 |
| 266.2786 | 270.5664 | 282.2196 |
| 292.5812 | 301.9914 | 305.2832 |
| 458.2187 | 474.2449 | 487.5152 |
| 491.5185 | 510.8966 | 516.4654 |
| 793.8405 | 838.9219 | 912.4641 |
| 927.6718 | 1019.5019 | 1043.5763 |
| 1151.7538 | 1179.6834 | 2577.5728 |
| 2592.5363 | 2683.1324 | 2717.3533 |